\documentclass[prl,twocolumn,amsmath,amssymb,floatfix,reprint,footinbib,superscriptaddress,longbibliography,showkeys]{revtex4-1}
\usepackage{dcolumn}
\usepackage{graphicx}
\usepackage{mathrsfs}
\usepackage{mdwlist}
\usepackage{subfigure}
\usepackage{booktabs}
\usepackage{amsmath}
\usepackage{dsfont}
\usepackage{amstext}
\usepackage{amssymb}
\usepackage{amsbsy}
\usepackage{bbm}
\usepackage{bm}
\usepackage{booktabs}
\usepackage{cases}
\usepackage{amsthm}
\usepackage{graphicx}
\usepackage{textcomp}
\usepackage{multirow}
\usepackage{color}
\usepackage{diagbox}
\usepackage[colorlinks,citecolor=blue]{hyperref}
\definecolor{Dgreen}{RGB}{0, 100, 0}
\usepackage{url}
\usepackage[colorlinks]{hyperref}
\hypersetup{%
	plainpages=true,
	breaklinks=true,  %not default in dvips mode, so we must specify
	hypertexnames=false,  %not ideal, but needed when pagenums duplicate (`i' vs. `1')
	pageanchor=true,
	colorlinks=true,
	linkcolor={blue},
	citecolor={blue},
	urlcolor={blue},
	anchorcolor={black}
}

\begin{document}
\title{Robust transitionless quantum driving: Concatenated approach}

\author{Zhi-Cheng Shi}
\affiliation{Fujian Key Laboratory of Quantum Information and Quantum Optics (Fuzhou University), Fuzhou
350108, China}
\affiliation{Department of Physics, Fuzhou University, Fuzhou 350108, China}

\author{Cheng Zhang}
\affiliation{Fujian Key Laboratory of Quantum Information and Quantum Optics (Fuzhou University), Fuzhou
350108, China}
\affiliation{Department of Physics, Fuzhou University, Fuzhou 350108, China}

\author{Li-Tuo Shen}
\affiliation{Fujian Key Laboratory of Quantum Information and Quantum Optics (Fuzhou University), Fuzhou
350108, China}
\affiliation{Department of Physics, Fuzhou University, Fuzhou 350108, China}

\author{Jie Song}
\affiliation{Department of Physics, Harbin Institute of Technology, Harbin 150001, China}

\author{Yan Xia}\thanks{xia-208@163.com}
\affiliation{Fujian Key Laboratory of Quantum Information and Quantum Optics (Fuzhou University), Fuzhou
350108, China}
\affiliation{Department of Physics, Fuzhou University, Fuzhou 350108, China}

\author{X. X. Yi} \thanks{yixx@nenu.edu.cn}
\affiliation{Center for Quantum Sciences and School of Physics, Northeast Normal University, Changchun 130024, China}

\begin{abstract}
We propose a concatenated approach for implementing transitionless quantum driving regardless of adiabatic conditions while being robustness with respect to all kinds of systematic errors induced by pulse duration, pulse amplitude, detunings, and Stark shift, etc. The current approach is particularly efficient for all time-dependent pulses with arbitrary shape, and only the phase differences between pulses is required to properly modulate.
The simple physical implementation without the help of pulse shaping techniques or extra pulses makes this approach quite universal and provides a different avenue for robust quantum control by the time-dependent Hamiltonian.

\end{abstract}

\maketitle

\emph{Introduction.} The adiabatic passage (AP) \cite{osti7365050,PhysRevA.29.690,RevModPhys.79.53} plays an important role in quantum optics \cite{scully97} and quantum computation \cite{Nielsen00,RevModPhys.76.1037,RevModPhys.90.015002}, due to the insensitivity to various errors in physical parameters.
In the adiabatic process,
the system always evolves along an eigenstate connecting the initial state with the target state.
One typical example of AP is the stimulated Raman adiabatic passage (STIRAP) \cite{PhysRevA.40.6741,PhysRevA.45.5297,RevModPhys.89.015006,PhysRevLett.126.113601,PhysRevLett.122.253201,PhysRevA.102.023515}, a popular way for population inversion in three-level systems.
Regardless of its robustness,
the common characteristic of AP are incompleteness of population transfer
and slow change of parameters over time.

To overcome these shortcomings,
a technique called shortcuts to adiabaticity (STA) \cite{Demirplak2003,Demirplak2005,Berry2009,PhysRevLett.105.123003,PhysRevLett.104.063002,PhysRevLett.111.100502, PhysRevLett.109.100403,PhysRevLett.122.173202,PhysRevLett.122.090502,PhysRevLett.122.050404,PhysRevLett.126.023602}
has emerged that can directly cancel nonadiabatic transitions by introducing extra counterdiabatic fields.
This technique allows the system to perfectly evolve along the eigenstate of the original Hamiltonian at a very fast rate but suffers a great loss on the robustness of AP since it requires exactly knowing the system parameters in advance. Recently, several works are devoted to the robustness of STA \cite{Ruschhaupt2012,PhysRevA.102.053104,PhysRevApplied.18.054055,liu2023shortcuts}.
Another essential requirement in STA is the pulse shaping technique to tailor the original waveform, and some counterdiabatic fields sometimes are prohibited from a physical point of view so its realization may be challenging.

As another alternative, the composite adiabatic passage (CAP) technique \cite{PhysRevLett.106.233001},
a combination of the best of both composite pulses \cite{Levitt1979,Wimperis1994,PhysRevA.67.042308,PhysRevLett.113.043001,PhysRevA.84.062311,Bando2013,PhysRevA.89.022310,PhysRevLett.118.150502,PhysRevLett.95.200501, PhysRevA.101.012321,PhysRevLett.129.240505,PhysRevResearch.2.043235,PhysRevA.104.012609,Shi2022,PhysRevApplied.18.034062} and adiabatic passage \cite{PhysRevA.96.013415,PhysRevA.101.013426}, has been proposed to achieve complete population inversion in two-level systems,
while its extension version named composite STIRAP \cite{PhysRevA.87.043418,PhysRevA.98.053413} achieves the same objective in three-level systems.
The main advantage of CAP over STA is the retained robustness of AP and the simplicity of implementation,
because one just properly modulates the relative phases between different pulses.
Nevertheless, some fundamental issues still need to be addressed in CAP, e.g., the unaccessible of obtaining universal rotation operations (or arbitrary superposition states) and the requirement of longer duration compared with traditional adiabatic passages.

In this work, through carefully designing the phase differences,
we propose a different dynamical mechanism for achieving perfect transitionless quantum driving in a robust manner, while simultaneously performing high precision quantum operations even \emph{without} knowing the magnitudes of various systematic errors.
The current approach is constructed by multiply concatenating the Hamiltonians with the same arbitrary pulse shape but different well-designed constant phases.
In particular, it is unnecessary to satisfy the adiabatic condition, and thus the total duration does not have to be long.

\begin{figure*}
\centering
\includegraphics[scale=0.545]{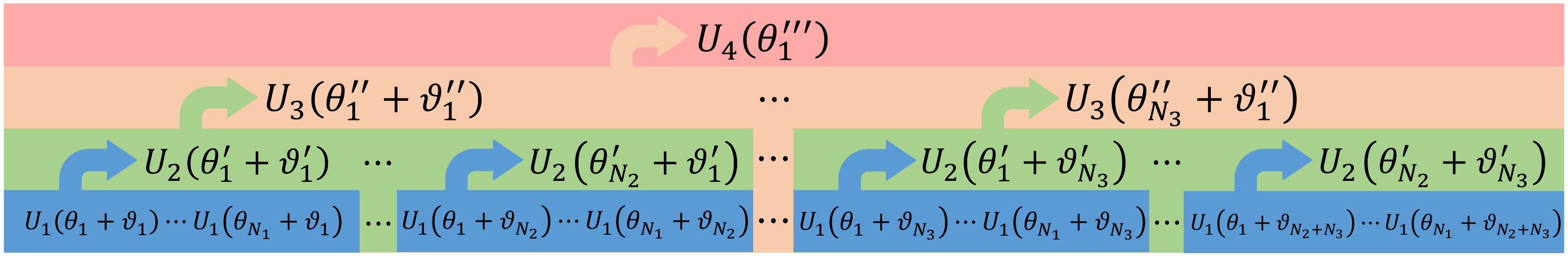}
\caption{Sketch diagram of the concatenated approach.
Different colors represent different uses, and $\bm{\vartheta}_1=\bm{\vartheta}_1'=\bm{\vartheta}_1''=0$. In the blue box (the first hierarchy), the phase differences $\bm{\theta}_{mn}$ in the propagator $U_2(\bm{\theta}_1')=U_1(\bm{\theta}_{N_1})\cdots U_1(\bm{\theta}_1)$ are applied to achieve perfect transitionless quantum driving. The only difference between the different blue boxes is the phase shift $\bm{\vartheta}_n$.
By continuing to concatenate the propagators $U_2(\bm{\theta}_n')$ in the green box  (the second hierarchy), i.e., $U_3(\bm{\theta}_1'')=U_2(\bm{\theta}_{N_2}')\cdots U_2(\bm{\theta}'_1)$, we design the phase differences $\bm{\theta}_{mn}'$ for the robustness against the nonadiabatic transition induced by various errors. Similarly, the phase differences $\bm{\theta}_{mn}''$ in the orange box (the third hierarchy) are devoted to performing high precision quantum operations, where $U_4(\bm{\theta}_1''')=U_3(\bm{\theta}_{N_3}'')\cdots U_3(\bm{\theta}''_1)$. Certainly, the propagators $U_4(\bm{\theta}_n''')$ can be further concatenated to realize other uses. In this way, we achieve transitionless quantum driving in a robust manner even in the presence of various systematic errors.}\label{sch}
\end{figure*}

\emph{Gauge invariance.} Consider a quantum system with near-neighbor interactions, and the general form of the time-dependent Hamiltonian is ($\hbar=1$)
\begin{eqnarray}
H(t)=\sum\lambda_{j,j}(t)|j\rangle\langle j|+\lambda_{j,j+1}(t)|j\rangle\langle j\!+\!1|+\mathrm{H.c.},~
\end{eqnarray}
where $\lambda_{j,j}(t)$ denote level energies, and $\lambda_{j,j+1}(t)$ are the coupling strengths of near-neighbor levels.
When introducing extra constant phases to the coupling strengths, the Hamiltonian becomes
\begin{eqnarray} \label{ham2}
H(t,\bm{\theta})&=&\sum\lambda_{j,j}(t)|j\rangle\langle j|+\lambda_{j,j+1}(t)e^{i\theta_{j}}|j\rangle\langle j\!+\!1|+\mathrm{H.c.} \cr
&=&\sum\lambda_{j,j}(t)|\tilde{j}\rangle\langle \tilde{j}|+\lambda_{j,j+1}(t)|\tilde{j}\rangle\langle \widetilde{j\!+\!1}|+\mathrm{H.c.},
\end{eqnarray}
where $\bm{\theta}=(\theta_{1}, \theta_{2}, \dots)$ represents a vector parametrizing different constant phases in external fields.
Through making appropriate transformations to the original basis: $|\tilde{j}\rangle=\exp(-{i\sum_{k=1}^{k=j-1}\theta_k})|j\rangle$, the form of the Hamiltonian with different constant phases is the same as the previous one.
Namely, the addition of constant phases simply means that we are choosing a set of new basis and the expression of the propagator remains unchanged, i.e., $U_1(t)=\tilde{U}_1(t,\bm{\theta})$, where $U_1(t)$ and $\tilde{U}_1(t,\bm{\theta})$ represent the propagators in the basis $|j\rangle$ and $|\tilde{j}\rangle$, respectively. Therefore, the physical property of the Hamiltonian~(\ref{ham2}) is still trivial after introducing arbitrary constant phases.
Actually, this trivial property originates from the gauge invariance of the system \cite{PhysRevA.93.052107}.

\emph{Nontriviality of phase differences}. Even though the trivial property do not change in any way for arbitrary constant phases,
\emph{phase differences} are of nontrivial physical significance and can be used for implementing perfect transitionless quantum driving in a robust manner.
To make it more clear,
we demonstrate the detailed design workflow in Fig.~\ref{sch}, where the propagator corresponding to the Hamiltonian $H(t,\bm{\theta}_n)$ reads $U_1(\bm{\theta}_n)=\mathbf{T}\exp{\left[-i\int H(t,\bm{\theta}_n)t\right]}$ with $\mathbf{T}$ being a time-ordering operator.
As shown in the blue box,
we initially concatenate $N_1$ Hamiltonians $H(t,\bm{\theta}_n)$ with the same time-varying shape but different $\bm{\theta}_n$ to create a composite propagator $U_2(\bm{\theta}_1')=U_1(\bm{\theta}_{N_1})\cdots U_1(\bm{\theta}_1)$, where
the phase differences $\bm{\theta}_{mn}=\bm{\theta}_m-\bm{\theta}_n$ are properly modulated to achieve perfect transitionless quantum driving.

When the quantum system exhibits various errors, transitionless quantum driving becomes imperfect and thus produces nonadiabatic transitions.
To solve it,
we continue to concatenating ${N_2}$ propagators $U_2(\bm{\theta}_{n}')$  to produce a new one $U_3(\bm{\theta}_1'')=U_2(\bm{\theta}_{N_2}')\cdots U_2(\bm{\theta}'_1)$, where $U_2(\bm{\theta}_{n}')$ are generated by adding different constant phase shifts $\bm{\vartheta}_n$ to $N_1$ Hamiltonians $H(t,\bm{\theta}_n)$.
Through altering the phase differences $\bm{\theta}'_{mn}=\bm{\theta}'_m-\bm{\theta}'_n$, the propagator $U_3(\bm{\theta}_1'')$ would sharply suppress the nonadiabatic transition induced by various errors.

Note that the propagator $U_3(\bm{\theta}_1'')$ becomes approximately diagonalized after the first two concatenations, but there still exists phase errors on diagonal elements and those errors would reduce the precision of quantum operations. Thus, we require to further concatenate ${N_3}$ propagators $U_3(\bm{\theta}_n'')$ to obtain the new one $U_4(\bm{\theta}_1''')=U_3(\bm{\theta}_{N_3}'')\cdots U_3(\bm{\theta}''_1)$.
Similarly, the phase differences $\bm{\theta}''_{mn}=\bm{\theta}''_m-\bm{\theta}''_n$ are finely tuned to promote the accuracy of quantum operations.
Indeed, the propagators $U_4(\bm{\theta}_n''')$ can be also concatenated for more other uses if necessary.
Through this concatenated approach, we obtain a sequence for implementing robust quantum operations under transitionless quantum driving.

\emph{Transitionless quantum driving.}
To illustrate this, let us consider a three-level $\Lambda$ system driven by a Stokes and pump (SP) pulse pair, while two-photon resonance is kept.
The form of the Hamiltonian reads
\begin{eqnarray}\label{hamiltonian}
\mathrm{H}(t)&\!=\!&\Delta(t)\sigma_{ee}\!+\!\left[\Omega_{p}(t)e^{i\theta_p}\sigma_{ge}\!+\!\Omega_{s}(t)e^{i\theta_s}\sigma_{fe} \!+\!\mathrm{H.c.}\right]\!, ~~~
\end{eqnarray}
where $\sigma_{kl}=|k\rangle\langle l|$ ($k,l=g,f,e$), and the $|g\rangle (|f\rangle) \leftrightarrow|e\rangle$ transition is driven by the pump (Stokes) pulse with the coupling strength $\Omega_{p}(t)$ [$\Omega_{s}(t)$], the phase $\theta_p$ ($\theta_s$), and the detuning $\Delta(t)$.
The instantaneous eigenstates of $H(t)$ are
$|d(t)\rangle=\cos\phi(t) \exp({-i\theta_{sp}})|g\rangle-\sin\phi(t)|f\rangle,~
|E_+(t)\rangle=\sin\varphi(t)\exp({i\theta_s})|b(t)\rangle +\cos\varphi(t)|e\rangle,~
\mathrm{and}~|E_-(t)\rangle=\cos\varphi(t)\exp{(i\theta_s)}|b(t)\rangle -\sin\varphi(t)|e\rangle$,
with $\theta_{sp}=\theta_s-\theta_p$,
$\Omega(t)=\sqrt{\Omega_{p}(t)^2+\Omega_{s}(t)^2}$,
$\tan2\varphi(t)=2\Omega(t)/\Delta(t)$,
$\tan\phi(t)=\Omega_{p}(t)/\Omega_{s}(t)$,
and the bright state $|b(t)\rangle=\sin{\phi(t)}|g\rangle+\cos{\phi(t)}\exp{(i\theta_{sp})}|f\rangle$.

In STIRAP, there is a time delay in the SP pulse pair.
Because of this asynchrony,
$\phi(t)$ is a variable quantity over time,
inevitably leading to nonadiabatic transitions between distinct eigenstates.
Here, we demand that the SP pulse pair must be synchronized [i.e., $\phi(t)$ is constant] to make the dark state $|d(t)\rangle$ completely decouple to other adiabatic states \cite{supp}.
To extract a freely adjustable phase (e.g., $\theta_s$), the eigenstates $|E_+(t)\rangle$ and $|E_-(t)\rangle$ are reassociated to form
a set of new basis $\{|d(t)\rangle, |b(t)\rangle, |e\rangle\}$, and we adopt this set of basis to reveal the dynamical mechanism of transitionless quantum driving.

In the synchronization of the SP pulse pair, the general expression for the system propagator in the basis $\{|d(t)\rangle, |b(t)\rangle, |e\rangle\}$ is given by
\begin{eqnarray}\label{propa}
U_1(\bm{\theta})&=&\left[
                \begin{array}{ccc}
                 1 & 0 & 0 \\[0.5ex]
                 0 & se^{i\alpha} & re^{-i\theta_s} \\[0.5ex]
                 0 & -r^*e^{i\theta_s} & se^{-i\alpha}
                \end{array}
              \right],
\end{eqnarray}
where $r$ and $\alpha$ are determined by the parameters of the SP pulse pair and $s=\sqrt{1-|r|^2}$.
Definitely, the appearance of $r$ in Eq.~(\ref{propa}) is the result of the coupling between the states $|b(t)\rangle$ and $|e\rangle$ in this system.
To completely nullify this nonadiabatic transition,
we just have to concatenate two SP pulse pairs with well-designed phase differences, and the propagator reads $U_2(\bm{\theta}_1')=U_1(\bm{\theta}_2)U_1(\bm{\theta}_1)$, where $U_1(\bm{\theta}_{n_1})$ is the propagator for the $n_1^{\mathrm{th}}$ SP pulse pair with $\bm{\theta}_{n_1}=(\theta_{s,n_1},\theta_{p,n_1})$ and $\theta_{s(p),{n_1}}$ refers to the phase of the $n_1^{\mathrm{th}}$ Stokes (pump) pulse, ${n_1}=1,2$.
When the phase differences satisfy \cite{supp}
\begin{eqnarray} \label{phasedif}
\!\!{\theta}_{21}=\theta_{s,2}-\theta_{s,1}=\pi-2\alpha,~~\theta_{s,n_1}-\theta_{p,n_1}=\theta_{sp},
\end{eqnarray}
the propagator becomes diagonalizable, i.e.,
\begin{eqnarray} \label{u6}
U_2(\bm{\theta}_1')&=&|d(t)\rangle \langle d(t)|\!+\!e^{i2\alpha}|b(t)\rangle \langle b(t)|\!+\!e^{-i2\alpha}|e\rangle \langle e|.~~~~
\end{eqnarray}
According to this propagator,
we can perfectly steer the system evolution along the dark/bright state without any transitions to the excited state $|e\rangle$.
Therefore, two SP pulse pairs are sufficient to achieve perfect transitionless quantum driving,
where the shape of the Stokes (pump) pulse can be arbitrary and does not require to satisfy the adiabatic condition.

When returning to the original representation, the propagator~(\ref{u6}) in the subspace spanned by $\{|g\rangle, |f\rangle\}$ becomes $\mathcal{U}=\exp\left({-i\alpha\bm{n}\cdot\bm{\sigma}}\right)$ with $\bm{\sigma}=(\sigma_x, \sigma_y, \sigma_z)$ being the Pauli operators. It actually represents a rotation operation around the axis $\bm{n}=(-\sin\phi(t)\cos\theta_{sp}, \sin\phi(t)\sin\theta_{sp}, \cos\phi(t))$ by the angle $\alpha$ in the Bloch sphere.
As a result,
we obtain a universal quantum gate, its evolution along the dark/bright state without any transitions.
In particular, different rotation operations [i.e., reflecting in different $\alpha$, $\phi(t)$, and $\theta_{sp}$] can be freely modulated by the coupling strength $\Omega_s(t)$, the detuning $\Delta(t)$, or the phase $\theta_s$ in this system.

\emph{Robustness.}
In reality, there are many factors that prevent us from exactly acquiring the full information of a quantum system.
For example, in the context of atoms driven by external fields, the external fields are usually assumed to be monochromatic while they actually have a certain linewidth. Due to their interactions, the energy level of atoms may also shift, i.e., the so-called Stark shift \cite{scully97}. Furthermore, the inhomogeneous distribution of external fields as well as tiny vibrations of atoms at equilibrium, can create small uncertainties in the interactions of these systems. All these uncertainties can be regarded as errors in various physical parameters, such as pulse duration, pulse amplitude, and detuning.
These errors mainly lead to two unfavorable effects: the generation of nonadiabatic transitions and the imprecision of quantum control.
Actually,
both effects can be largely avoided by further concatenating the propagators with well-designed constant phases.
Next, we elaborate on this point.

\begin{figure*}
\centering
\includegraphics[scale=0.62]{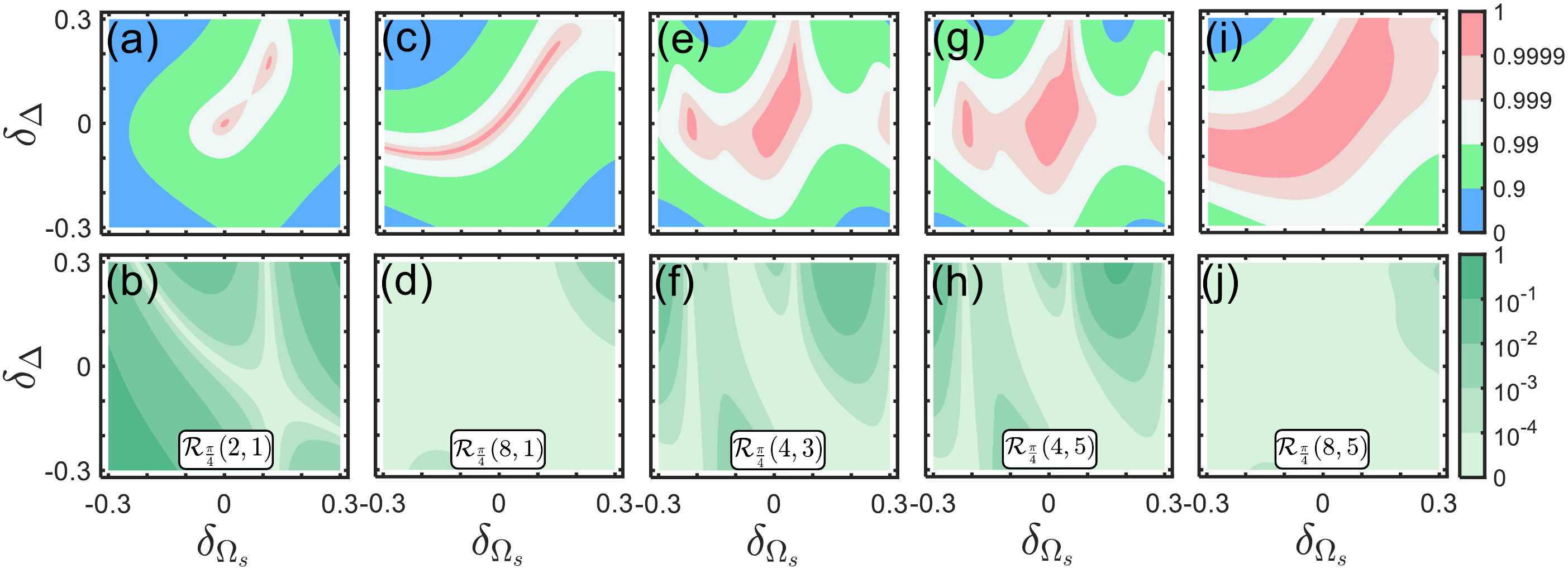}
\caption{Fidelity $F$ of the target state $|\Psi_T\rangle$ (top panels) and population $P_e$ of the state $|e\rangle$ (bottom panels) vs pulse amplitude error $\delta_{\Omega_s}$ and detuning error $\delta_\Delta$ for different sequences.
The fidelity is defined by $F=|\langle\Psi_T|\Psi(t)\rangle|^2$ and $\mathcal{R}_\alpha(N_1\times N_2,N_3)$ refers to performing a rotation operation around the axis $\bm{n}$  by the angle $\alpha$ in the Bloch sphere to obtain the desired state $\cos\alpha|g\rangle+\sin\alpha|f\rangle$, where $|\Psi_T\rangle=1/\sqrt{2}(|g\rangle+|f\rangle)$, the initial state is $|g\rangle$, and the total number of SP pulse pairs is $N_1\times N_2\times N_3$ with $N_1=2$. The $n^{\mathrm{th}}$ Stokes pulse has a Gaussian shape $\Omega_s(t)=A\Omega_0 \exp[(t-3n\tau)^2/\tau^2]$ with the duration $T=6\tau$. }\label{fig3}
\end{figure*}

We can see in Eq.~(\ref{propa}) that systematic errors mainly cause the deviation of two quantities: $r$ and $\alpha$.
Fortunately,
the concatenated dynamical mechanism for transitionless quantum driving is inherently immune to the deviation in the quantity $r$,
because the design of phase differences is independent of $r$ \cite{supp}.
As a result,
the system evolution is completely unaffected by the deviation in the quantity $r$.
For the deviation in the quantity $\alpha$,
it makes the phase difference satisfying Eq.~(\ref{phasedif}) slightly different,
resulting in the generation of the nonadiabatic transition.
To reduce this transition,
we can concatenate $N_2$ propagators with different constant phases: $U_3({\theta}_1'')=U_2(\bm{\theta}_{N_2}')\cdots U_2(\bm{\theta}'_1)$, where $\bm{\theta}_{n_2}'=(\theta_{s,{n_2}},\theta_{p,{n_2}})$ and $\theta_{s(p),{n_2}}$ refers to the phase of the $(2n_2-1)^{\mathrm{th}}$ Stokes (pump) pulse in this situation, $n_2=1,\dots,N_2$.
For $N_2=2^{M}$, $M=1,2,\dots$, when the phase differences satisfy \cite{supp}
\begin{eqnarray}
\theta'_{2^M n_2-2^{M-1}+1,2^M n_2-2^M+1}=\pi-2^{M+1}\alpha,
\end{eqnarray}
where $\theta_{m,n}'=\theta_{s,m}-\theta_{s,n}$,
the sequence is accurate to the $(M+1)^{\mathrm{th}}$ order deviation. For other pulse numbers, i.e., $N_2\neq2^{M}$, we can adopt the concatenated method or numerical method to obtain the phase differences \cite{supp}.

On the nonadiabatic transition induced by the deviation in the quantity $\alpha$ being dramatically suppressed,
the propagator $U_3({\theta}_1'')$ can be approximately written as a diagonalizable form: $U_3({\theta}''_1)\approx|d(t)\rangle \langle d(t)|+\exp[{i2\beta(1+\delta_{\beta})}]|b(t)\rangle \langle b(t)|+\exp[{-i2\beta'(1+\delta_{\beta'})}]|e\rangle \langle e|$.
It is worth mentioning that the deviation $\delta_{\beta}$ cannot be completely eliminated in the second concatenation \cite{Wang2012} and thus introduces the fault of quantum operations.
To decline its influence,
we need to return back to the original basis $\{|g\rangle, |f\rangle\}$
and then further concatenate $N_3$ propagators with distinct phase differences:
$U_4({\theta}_1''')=U_3({\theta}_{N_3}'')\cdots U_3({\theta}''_1)$,
where $U_3({\theta}''_{n_3})\approx\exp\left[{-{i}\beta(1+\delta_{\beta})\bm{n}\cdot\bm{\sigma}}\right]$ and ${\theta}_{n_3}''$ refers to the phase differences $\theta_{sp,{n_3}}$ between the $(4n_3-1)^{\mathrm{th}}$ Stokes pulse and pump pulse, $n_3=1,\dots,N_3$.
Here,
the pulse number $N_3(N_3\geq3)$ can be arbitrarily selected.
When $N_3$ is small, e.g., $N_3=3$,
the analytical expression of phase differences is given by \cite{supp}
\begin{subequations} \label{8probility}
\begin{eqnarray}
\theta_{21}''&=&2\arctan \pm\left(\sqrt{1-P_f^2}\pm\sqrt{2P_f-P_f^2}\right), \\
\theta_{32}''&=&2\arctan \pm\left(\sqrt{1-P_f^2}\mp\sqrt{2P_f-P_f^2}\right),
\end{eqnarray}
\end{subequations}
where $\theta_{mn}''=\theta_{m}''-\theta_{n}''$ and $P_f$ represents the population of the state $|f\rangle$.
As for a longer sequence (i.e., $N_3>3$),
the performance of the robustness against the deviation $\delta_{\beta}$ becomes much better since more adjustable phases are contained, and it is instructive to adopt numerical calculations to obtain the solutions.

Note that the phase differences given by Eq.~(\ref{8probility}) are used to compensate for the population deviation of the target state $|\psi_T\rangle$. To be able to simultaneously compensate for the phase deviation of the target state, we require to perform a Taylor expansion of the fidelity $F=|\langle\Psi_T|\Psi(N_3T)\rangle|^2$ instead, where $|\Psi(N_3T)\rangle$ represents the final state after concatenating $N_3$ pulses.
Definitely, the process of resolving phase differences is similar to that of using population formulas (see also in Ref.~\cite{supp}), and such a design actually helps to suppress the phase sensitivity of the target state. Furthermore, with the same number of pulses the system robustness designed by fidelity formulas may be marginally inferior to that of using population formulas. The reason is obvious, i.e., more phases are involved to compensate for the deviation $\delta_{\beta}$. Therefore, a longer sequence may be necessary to achieve the same robust effect in this situation.

In Fig.~\ref{fig3},
we demonstrate the robust performance of the quantum operation in the presence of two systematic errors
by different $\mathcal{R}_{\frac{\pi}{4}}(N_1\times N_2,N_3)$ sequences, where $\mathcal{R}_{\frac{\pi}{4}}(N_1\times N_2,N_3)$ represents the rotation around the axis $\bm{n}$ with the angle ${{\pi}/{4}}$, and the total pulse number $N=N_1\times N_2\times N_3$ with $N_1=2$.
For simplicity, the shape of each Stokes pulse is chosen as Guassian here, while other shapes can still work well \cite{supp}.

As shown in Figs.~\ref{fig3}(a) and \ref{fig3}(b), a concatenation of two SP pulse pairs to achieve perfect transitionless quantum driving possesses preliminary robustness against the pulse amplitude and detuning errors, whereas it does not work very well on large systematic errors.
When the SP pulse pairs are continuously recombined for a second time [cf. $N_2=4$ in Figs.~\ref{fig3}(c) and \ref{fig3}(d)],
the robust performance of transitionless quantum driving is dramatically enhanced so as to obtain an extremely low leakage population of the state $|e\rangle$.
Note that
this recombination makes little contribution to promoting the precision of rotation operations,
since the fidelity is sensitive to errors yet.
Therefore, we need to execute the third concatenation of SP pulse pairs.
Figures~\ref{fig3}(e)--\ref{fig3}(j) shows that
the high-fidelity region gradually enlarges as $N_3$ increases,
while the ability of leakage suppression is still reserved, implying that the deviation $\delta_{\beta}$ can be favorably compensated by properly designing the phase difference of each pulse pair.
Certainly, when concurrently increasing $N_2$ and $N_3$,
both population leakage and operation precision are efficiently improved.

\begin{figure}[t]
	\centering
    \includegraphics[scale=0.35]{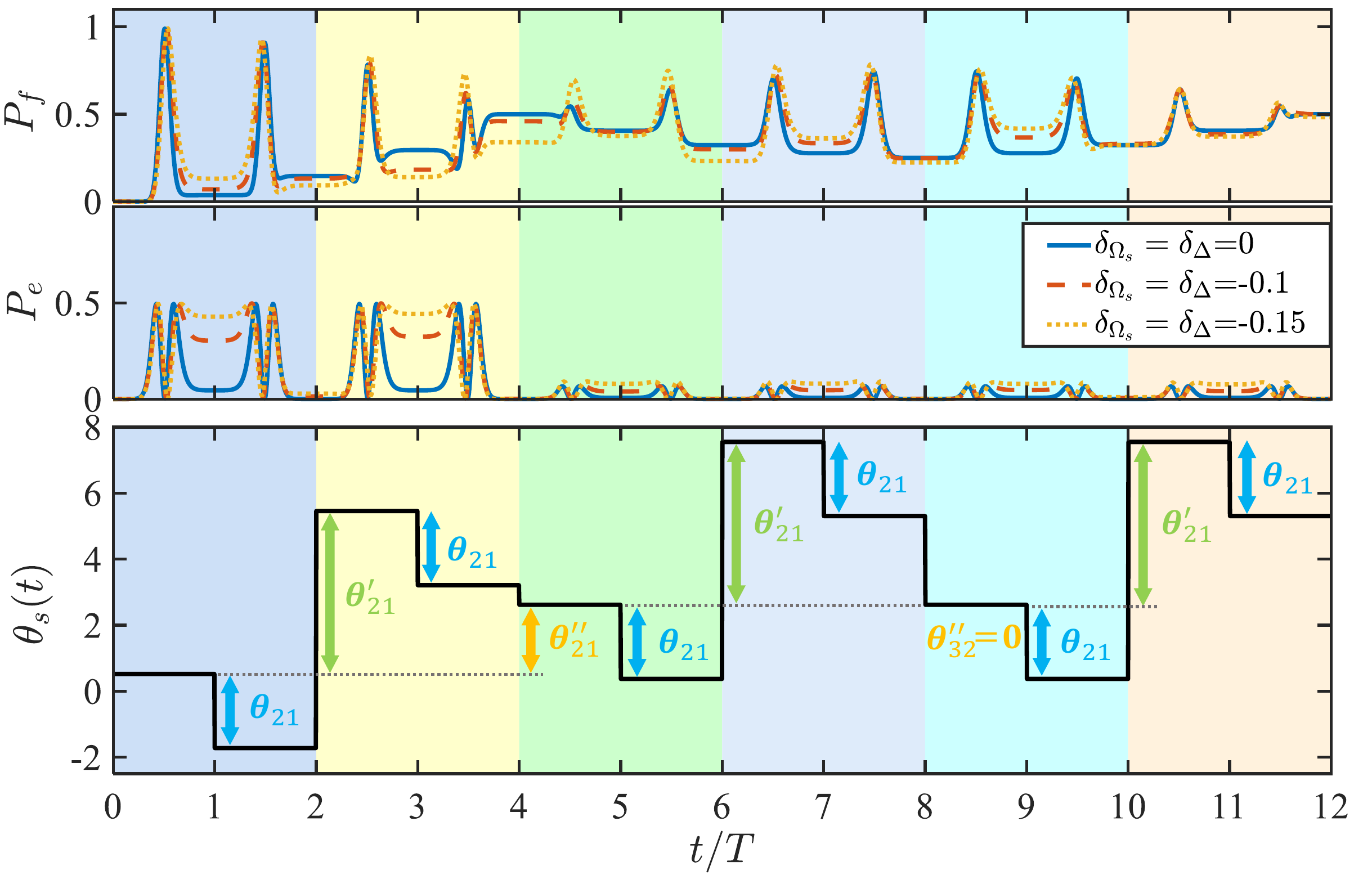}
	\caption{
Robust generation of the state $|\Psi_T\rangle$ and phase waveform of the Stokes pulse for the $\mathcal{R}_{\frac{\pi}{4}}(4,3)$ sequence.
The top (middle) panel represents the population evolution of the state $|f\rangle$ ($|e\rangle$) in the absence/presence of the errors $\delta_{\Omega_s}$ and $\delta_{\Delta}$.  After 12 SP pulse pairs with well-designed phase differences, the system is primely driven to the state $|\Psi_T\rangle$ even though it exhibits significant pulse amplitude and detuning errors; see the dashed and dotted curves.
In the bottom panel, the phase differences $\theta_{21}$ between the $2n^{\mathrm{th}}$ and $(2n-1)^{\mathrm{th}}$ SP pulse pairs are used for perfect transitionless quantum driving. The phase differences $\theta_{21}'$ between the $(4n-1)^{\mathrm{th}}$ and $(4n-3)^{\mathrm{th}}$ SP pulse pairs are devoted to suppress the nonadiabatic transition induced by the pulse amplitude and detuning errors. The phase differences $\theta_{(n+1)n}''$ between the $(4n+1)^{\mathrm{th}}$ and $(4n-3)^{\mathrm{th}}$ SP pulse pairs are employed for improving the fidelity of the Hadamard gate with a specific phase.
All phase differences $\theta_{mn}$, $\theta_{mn}'$, and $\theta_{mn}''$ are accessible by properly modulating the phases $\theta_s$ and $\theta_p$ of SP pulse pairs.}\label{2Npopulation2}
\end{figure}

To see more clearly, we demonstrate in Fig.~\ref{2Npopulation2} the population evolution of two states $|f\rangle$ and $|e\rangle$ and the phase waveform of the Stokes pulse by the $\mathcal{R}_{\frac{\pi}{4}}(4,3)$ sequence.
For a single SP pulse pair,
the system exhibits nonadiabatic transitions, i.e., the leakage to the excited state $|e\rangle$ (see the blue curve at $t=T$), because the adiabatic condition is broken.
By concatenating two SP pulse pairs and properly adjusting the phase difference $\theta_{21}$,
the transition to the excited state is completely eliminated, as shown by the blue curve at $t=2T$.
When the systematic errors are large,
the evolution path seriously deviates from the original one, as shown by the dashed and dotted curves.
We then concatenate four SP pulse pairs and modulate the phase difference $\theta_{21}'$ to reduce the unfavorable influence on the transition of the excited state induced by two systematic errors.
In this circumstance, the nonadiabatic transition is sharply suppressed,
but the fidelity of the rotation operation is not much improved; see the curves at $t=4T$.
Through further concatenating three groups of four SP pulse pairs (12 in total) and controlling the phase difference $\theta_{21}''$ and $\theta_{32}''$,
the system evolution is strictly restricted in the subspace $\{|g\rangle, |f\rangle\}$,
and the rotation operation becomes remarkable error tolerant.

\emph{Discussion.}
The concatenated approach may be applicable to various kinds of quantum systems as long as phase modulation is accessible, although we just take the typical three-level system to illustrate this issue. Actually,  the
phases only have to satisfy some constraints rather than arbitrary values when the Hamiltonian has long-ranged interactions \cite{supp}.
On the other hand, there are several basic requirements in the current STA (e.g, see a recent review \cite{RevModPhys.91.045001} and the references therein), such as multiparameter regulation, pulse shaping, and specific operation time. Also, the STA technique makes it difficult to make the system simultaneously inhibit multiple systematic errors.
Nevertheless, none of them are required in the concatenated approach, since the dynamical procedure is completely out of line with the
original framework of STA.
In particular, the design of phase differences does not rely on the certain type of systematic errors, because they are derived from the propagator instead of the Hamiltonian. Therefore, this approach is quite universal for resisting any errors.

\emph{Conclusion.}
We have developed a concatenated approach of constructing a dynamical mechanism for both achieving perfect transitionless quantum driving and improving the robustness with respect to various systematic errors.
It is particularly significant that the pulse shape can be arbitrary while the quantum system does not have to satisfy the adiabatic condition.
By properly designing the phase differences between different Hamiltonians, the unwanted nonadiabatic transitions
are sharply suppressed even in the presence of all kinds of errors. Meanwhile, quantum operations with very high fidelity are naturally obtained. Of course, this concatenated procedure can still be carried on for other uses.
The simplicity, flexibility, and versatility of the concatenated approach opens a promising avenue for high precision control in quantum information processing.

\begin{acknowledgments}
This work is supported by the Natural Science Foundation of Fujian Province under
Grant No. 2021J01575, the Natural Science Funds
for Distinguished Young Scholar of Fujian Province
under Grant No. 2020J06011, and the Project from Fuzhou
University under Grant No. JG202001-2.
\end{acknowledgments}

\bibliographystyle{apsrev4-1}
\bibliography{references}

\clearpage
\begin{widetext}

\setcounter{equation}{0}
\renewcommand\theequation{S\arabic{equation}}

\setcounter{figure}{0}
\renewcommand\thefigure{S\arabic{figure}}

\setcounter{section}{0}
\renewcommand\thesection{S\arabic{section}}

\setcounter{table}{0}
\renewcommand\thetable{S\Roman{table}}

\subsection{\Large{Supplementary Material for ``Robust transitionless quantum driving: Concatenated approach''}}

\vspace{2em}

\section{SI. Dark state decoupling from other eigenstates}\label{appda}

In this section,
we give a detailed derivation process that the Stokes and pump (SP) pulse pair must be synchronized for the dark state decoupling from other eigenstates. The system of interest is a three-level $\Lambda$ structure and driven by a Stokes and pump (SP) pulse pair under the two-photon resonance condition.
The form of the Hamiltonian reads ($\hbar=1$)
\begin{eqnarray}
\mathrm{H}(t)&=&\Delta(t)|e\rangle\langle e|+\Omega_{p}(t)e^{i\theta_p}|g\rangle\langle e|+\Omega_{s}(t)e^{i\theta_s}|f\rangle\langle e|+\mathrm{H.c.},
\end{eqnarray}
where the transition $|g\rangle (|f\rangle) \leftrightarrow|e\rangle$ is driven by the pump (Stokes) pulse with the coupling strength $\Omega_{p}(t)$ [$\Omega_{s}(t)$], the phase $\theta_p$ ($\theta_s$), and the detuning $\Delta(t)$.
For this Hamiltonian, we easily obtain the eigenenergies
\begin{subequations}
\begin{eqnarray}
E_0(t)&=&0,  \\[0.5ex]
E_+(t)&=&\frac{\Delta(t)}{2}+\sqrt{\frac{\Delta(t)^2}{4}+\Omega(t)^2}=\Omega(t)\cot\varphi(t), \\[0.5ex] E_-(t)&=&\frac{\Delta(t)}{2}-\sqrt{\frac{\Delta(t)^2}{4}+\Omega(t)^2}=-\Omega(t)\tan\varphi(t),
\end{eqnarray}
\end{subequations}
and the corresponding eigenstates
\begin{subequations}
\begin{eqnarray}
|E_0(t)\rangle&=&|d(t)\rangle=\cos\phi(t)e^{-i\theta_{sp}} |g\rangle-\sin\phi(t)|f\rangle, \label{eigenstate1}\\[0.5ex]
|E_+(t)\rangle&=&\sin\phi(t)\sin\varphi(t)e^{i\theta_p}|g\rangle+ \cos\phi(t)\sin\varphi(t) e^{i\theta_s}|f\rangle+\cos\varphi(t)|e\rangle
=\sin\varphi(t)e^{i\theta_s}|b(t)\rangle +\cos\varphi(t)|e\rangle, ~~~~~\label{eigenstate2}\\[0.5ex]
|E_-(t)\rangle&=&\sin\phi(t)\cos\varphi(t)e^{i\theta_p}|g\rangle+ \cos\phi(t)\cos\varphi(t) e^{i\theta_s}|f\rangle-\sin\varphi(t)|e\rangle
=\cos\varphi(t)e^{i\theta_s}|b(t)\rangle -\sin\varphi(t)|e\rangle,\label{eigenstate3}
\end{eqnarray}
\end{subequations}
where $\Omega(t)=\sqrt{\Omega_{p}(t)^2\!+\!\Omega_{s}(t)^2}$,
$\tan2\varphi(t)=2\Omega(t)/\Delta(t)$,
$\tan\phi(t)=\Omega_{p}(t)/\Omega_{s}(t)$, $\theta_{sp}=\theta_s\!-\!\theta_p$ is the phase difference between the Stokes and pump pulse,
and $|b(t)\rangle=\sin{\phi(t)}|g\rangle+\cos{\phi(t)}e^{i\theta_{sp}}|f\rangle$ represents the bright state.

The transition mechanism of the dark state is very easily revealed in the adiabatic representation spanned by Eqs.~(\ref{eigenstate1})--(\ref{eigenstate3}).
The matrix form of the Hamiltonian in the adiabatic basis $\{|E_0(t)\rangle, |E_+(t)\rangle,|E_-(t)\rangle\}$ reads
\begin{eqnarray}\label{hamiltonian111}
H_a(t)=\left[
                \begin{array}{ccc}
                                                       0&         -i\dot{\phi}(t)\sin{\varphi(t)}&      -i\dot{\phi}(t)\cos{\varphi(t)}\\
         i\dot{\phi}(t)\sin{\varphi(t)}&               \Omega(t)\cot{\varphi(t)}&        i\dot{\varphi}(t)  \\
        i\dot{\phi}(t)\cos{\varphi(t)}&                            -i\dot{\varphi}(t) &       -\Omega(t)\tan{\varphi(t)}
                \end{array}
              \right],
\end{eqnarray}
where the overdot represents a time derivative.
Obviously,
to make the dark state $|d(t)\rangle$ decouple to other adiabatic states,
we have to set $\dot{\phi}(t)=0$,
and the solution is
\begin{eqnarray}
{\phi}(t)=\mathrm{constant}.
\end{eqnarray}
This means the SP pulses must be synchronized.
In fact,
this synchronization between the SP pulses over time is an essential condition to make one state decouple to other states.
In the following, we give a brief proof for it.

Consider a general unitary transform $R$ on this system, and the matrix form of $R$ in the basis $\{|g\rangle, |f\rangle, |e\rangle\}$ is
\begin{eqnarray}
R(t)&=&\left[
                \begin{array}{ccc}
             \cos{\phi(t)}e^{-i\vartheta} & \sin{\phi(t)}&0  \\[1.0ex]
                -\sin{\phi(t)}& \cos{\phi(t)}e^{i\vartheta}& 0\\[1.0ex]
                0&0&1\cr
                \end{array}
              \right].
\end{eqnarray}
We now study the system dynamics in a rotation frame with respect to $R$, where the basis are $\{|r_1(t)\rangle, |r_2(t)\rangle, |e\rangle\}$ with $|r_1(t)\rangle=\cos{\phi(t)}e^{-i\vartheta}|g\rangle-\sin{\phi(t)}|f\rangle$ and
$|r_2(t)\rangle=\sin{\phi(t)}|g\rangle+\cos{\phi(t)}e^{i\vartheta}|f\rangle$.
Here, the undetermined parameters $\phi(t)$ and $\vartheta$ are used to ensure that one basis is decoupled from other two bases in this rotation frame.
We first calculate the expression of the Hamiltonian in the basis $\{|r_1(t)\rangle, |r_2(t)\rangle, |e\rangle\}$, i.e.,
\begin{eqnarray}
H_r(t)&=&R^{\dagger}(t)H(t)R(t)-iR^{\dagger}(t)\dot{R}(t),\nonumber\\[0.5ex]
&=&\left[
                \begin{array}{ccc}
                0 & h_{12}&h_{13}  \\[1.0ex]
               h_{21}& 0& h_{23}\\[1.0ex]
                h_{31}&h_{32}&h_{33}\cr
                \end{array}
              \right]
\end{eqnarray}
where
\begin{subequations}
\begin{eqnarray}
h_{12}&=&h_{21}^*=-i\dot{\phi}(t)e^{-i\vartheta}, \\[1ex]
h_{13}&=&h_{31}^*=\Omega_{p}(t)\cos\phi(t)e^{i(\theta_p+\vartheta)}-\Omega_{s}(t)\sin\phi(t)e^{i\theta_s}, \\[1ex]
h_{23}&=&h_{32}^*=\Omega_{p}(t)\sin\phi(t)e^{i\theta_p}+\Omega_{s}(t)\cos\phi(t)e^{i(\theta_s-\vartheta)}, \\[1ex]
h_{33}&=&\Delta(t).
\end{eqnarray}
\end{subequations}
It can be found that
it is required $h_{12}=h_{13}=0$ to decouple $|r_1(t)\rangle$ from $|r_2(t)\rangle$ and $|e\rangle$,
or $h_{12}=h_{32}=0$ to decouple $|r_2(t)\rangle$ from $|r_1(t)\rangle$ and $|e\rangle$.
For $h_{12}=0$, we have $\dot{\phi}(t)=0$, and the solution is
\begin{eqnarray}
{\phi}(t)=\mathrm{constant}.
\end{eqnarray}
On the other hand,
the equation $h_{13}=0$ can be simplified as
\begin{eqnarray}
\tan\phi(t)=\frac{\Omega_{p}(t)}{\Omega_{s}(t)}e^{i(\theta_p-\theta_s+\vartheta)},
\end{eqnarray}
while the equation $h_{23}=0$ is simplified as
\begin{eqnarray}
\cot\phi(t)=-\frac{\Omega_{p}(t)}{\Omega_{s}(t)}e^{i(\theta_p-\theta_s+\vartheta)}.
\end{eqnarray}
As a result, we acquire that $\vartheta=\theta_s-\theta_p=\theta_{sp}$ and the Stokes and pump pulses must be synchronized, since $\phi(t)$ is real constant.

\section{Perfect transitionless quantum driving by two SP pulse pairs}

In the section, we demonstrate that perfect transitionless quantum driving is obtained by merely concatenating two SP pulse pairs with a well-designed phase difference.

Due to the decoupling between the dark state $|b(t)\rangle$ and the states $|b(t)\rangle$ and $|e\rangle$,
the general expression of the propagator for the single SP pulse pair in the basis $\{|d(t)\rangle, |b(t)\rangle, |e\rangle\}$ is written as
\begin{eqnarray}\label{20propa}
U_1(\bm{\theta})&=&\left[
                \begin{array}{ccc}
                 1 & 0 & 0 \\[0.5ex]
                 0 & se^{i\alpha} & re^{-i\theta_s} \\[0.5ex]
                 0 & -r^*e^{i\theta_s} & se^{-i\alpha}
                \end{array}
              \right],
\end{eqnarray}
where $r$ and $\alpha$ are determined by the parameters of the SP pulse pair and $s=\sqrt{1-|r|^2}$.
Thus, the propagator for two SP pulse pairs is given by
\begin{eqnarray}\label{21propa}
U_2(\bm{\theta}_1')=U_1(\bm{\theta}_2)U_1(\bm{\theta}_1)=\left[
\begin{array}{ccc}
                 1 & 0          &  0  \\[1ex]
                 0                    &{U}_{2,bb}&{U}_{2,be} \\[1ex]
                 0                    &{U}_{2,eb}&{U}_{2,ee}
\end{array}
\right],\!
\end{eqnarray}
where
\begin{subequations}
\begin{eqnarray}
{U}_{2,bb}&=&{U}^*_{2,ee}=-|r|^2 e^{i(\theta_{s,1}-\theta_{s,2})} +s^2e^{i2\alpha},\\
{U}_{2,eb}&=&-{U}^*_{2,be}=-r^* s \left[e^{-i(\alpha-\theta_{s,1})}+e^{i(\alpha+\theta_{s,2})}\right],
\end{eqnarray}
\end{subequations}
and $\theta_{s,1}$ ($\theta_{s,2}$) are the phase of the first (second) Stokes pulse.
To perfectly implement transitionless quantum driving,
we need to nullify the matrix element ${U}_{2,eb}$, and thus the equation becomes
\begin{eqnarray} \label{55phase}
e^{-i(\alpha-\theta_{s,1})}+e^{i(\alpha+\theta_{s,2})}=0.
\end{eqnarray}
It is easily to obtain the solution of Eq.~(\ref{55phase}), i.e.,
\begin{eqnarray}\label{phase2}
\theta_{s,2}-\theta_{s,1}=\pi-2\alpha.
\end{eqnarray}
This is actually Eq.~(5) in the main text.
As a result, the propagator given by Eq.~(\ref{21propa}) becomes diagonalized, i.e.,
\begin{eqnarray}
U_2(\bm{\theta}_1')=\left[
\begin{array}{ccc}
                 1 & 0          &  0  \\[0.5ex]
                 0                    &e^{i2\alpha}&0  \\[0.5ex]
                 0                    &0&e^{-i2\alpha}
\end{array}
\right].\!~~~
\end{eqnarray}
Therefore, we can perform perfect transitionless quantum driving by only two SP pulse pairs regardless of whether it exists the nonadiabatic transition in the single SP pulse pair.

\section{Robustness against nonadiabatic transition induced by various systematic errors} \label{siii}

In this section, we demonstrate how to concatenate $N_2$ propagators $U_2(\bm{\theta}_{n_2}')$ with different constant phases to achieve robust transitionless quantum driving in the presence of systematic errors, i.e., the second concatenation in the green box of Fig.~1 in the main text.

When the quantum system exhibits various errors (e.g., pulse amplitude error, pulse duration error, and detuning error etc.), it leads to the deviation of two quantities $r$ and $\alpha$ in the propagator $U_2(\bm{\theta}_{n_2}')$.
In the following, let us study this issue separately.

(1) \emph{The deviation of the quantities} $r$ ($s$): $r\rightarrow (1+\delta_r)r$ [$s\rightarrow (1+\delta_s)s$].
The propagator $U_1(\bm{\theta}_1)$ for the single SP pulse pair becomes
\begin{eqnarray}  \label{26pro}
U_1(\bm{\theta}_1)&=&\left[
                \begin{array}{ccc}
                 1 & 0 & 0 \\[0.5ex]
                 0 & (1+\delta_s)se^{i\alpha} & (1+\delta_r)re^{-i\theta_{s,1}} \\[0.5ex]
                 0 & -(1+\delta_r)r^*e^{i\theta_{s,1}} & (1+\delta_s)se^{-i\alpha}
                \end{array}
              \right],
\end{eqnarray}
Then, the propagator for two SP pulse pairs is given by
\begin{eqnarray}\label{2pulsepropa}
U_2(\bm{\theta}_1')=U_1(\bm{\theta}_2)U_1(\bm{\theta}_1)=\left[
\begin{array}{ccc}
                 1 & 0          &  0  \\[1ex]
                 0                    &{U}_{2,bb}&{U}_{2,be} \\[1ex]
                 0                    &{U}_{2,eb}&{U}_{2,ee}
\end{array}
\right],
\end{eqnarray}
where
\begin{subequations}
\begin{eqnarray}
{U}_{2,bb}&=&{U}^*_{2,ee}=-(1+\delta_r)^2|r|^2 e^{i(\theta_{s,1}-\theta_{s,2})} +(1+\delta_s)^2s^2e^{i2\alpha},\\
{U}_{2,eb}&=&-{U}^*_{2,be}=-(1+\delta_r)(1+\delta_s)r^* s \left[e^{-i(\alpha-\theta_{s,1})}+e^{i(\alpha+\theta_{s,2})}\right].\label{s20b}
\end{eqnarray}
\end{subequations}
Obviously, the solution of Eq.~(\ref{s20b}) is
\begin{eqnarray} \label{s28}
\theta_{s,2}-\theta_{s,1}=\pi-2\alpha,
\end{eqnarray}
which is independent of the deviation in the quantity $r$ ($s$).
Therefore, once the phase difference satisfy Eq.~(\ref{s28}) in the two SP pulse pairs, one can completely eliminate the effect of the deviation in the quantity $r$ ($s$) no matter how large it is.

(2) \emph{The deviation of the quantity $\alpha$}: $\alpha\rightarrow (1+\delta_\alpha)\alpha$.
In this case,
the propagator $U_1(\bm{\theta}_1)$ becomes
\begin{eqnarray}\label{28propagator}
U_1(\bm{\theta}_1)=\left[
                \begin{array}{ccc}
                 1 & 0 & 0 \\[1ex]
                 0 & se^{i(1+\delta_{\alpha})\alpha} & re^{-i\theta_{s,1}} \\[1ex]
                 0 & -r^*e^{i\theta_{s,1}} & se^{-i(1+\delta_{\alpha})\alpha}
                \end{array}
              \right].
\end{eqnarray}
As a result, the propagator $U_2(\bm{\theta}_1')$ is
\begin{eqnarray} \label{29pro}
U_2(\bm{\theta}_1')=U_1(\bm{\theta}_2)U_1(\bm{\theta}_1)=\left[
\begin{array}{ccc}
                 1 & 0          &  0  \\[1ex]
                 0                    &e^{2i\alpha}(|r|^2+e^{2i\alpha\delta_\alpha}s^2)& 2irs\sin(\alpha\delta_\alpha)e^{i(\alpha-\theta_{s,1})} \\[1ex]
                 0                    &2ir^*s\sin(\alpha\delta_\alpha)e^{i(\theta_{s,1}-\alpha)}&e^{-2i\alpha}(|r|^2+e^{-2i\alpha\delta_\alpha}s^2)
\end{array}
\right].
\end{eqnarray}
Obviously, Eq.~(\ref{29pro}) demonstrates that the nonadiabatic transition is induced by the presence of the deviation $\delta_\alpha$.

It is worth mentioning that the unfavorable effect caused by the deviation $\delta_\alpha$ can only be sharply reduced rather than completely canceled in this situation.
Next, we elaborate the implementation of robust transitionless quantum driving by designing the phase differences in a train of $N_2$ propagators $U_2(\bm{\theta}_{n_2}')$,
where the recombination propagator $U_3({\theta}_1'')$ reads
\begin{eqnarray}  \label{s31}
U_3({\theta}_1'')=U_2(\bm{\theta}_{N_2}')\cdots U_2(\bm{\theta}'_1)=\left[
\begin{array}{ccc}
                 1 & 0          &  0  \\[1ex]
                 0                    &{U}_{3,bb}&{U}_{3,be} \\[1ex]
                 0                    &{U}_{3,eb}&{U}_{3,ee}
\end{array}
\right],
\end{eqnarray}
with $\bm{\theta}_{n_2}'=(\theta_{s,{n_2}},\theta_{p,{n_2}})$, $n_2=1,\dots,N_2$.
Through the Tayor-Maclaurin expansion for the propagator (\ref{s31}) with respect to $\delta_\alpha$, we can obtain the series of the element ${U}_{3,eb}$.
Then, to achieve robustness against the deviation $\delta_\alpha$, we require to design the phase differences to nullify the derivatives in a increasing order as many as possible, i.e.,
\begin{eqnarray}\label{s32}
\partial^{k}_{\delta_\alpha}\!U_{3,eb}=0,~~~ k=0,1,2,3,\dots
\end{eqnarray}
where $\partial^{k}_{\delta_\alpha}\!U_{3,eb}$ represents the $k^\mathrm{th}$ derivative of $U_{3,eb}$ with respect to $\delta_\alpha$, and $k=0$ corresponds to the zero-error value.
When the number of the propagators $U_2(\bm{\theta}_{n_2}')$ is small, we can analytically obtain the solution of Eq.~(\ref{s32}).
For a longer sequence, the solution can be given by concatenating the obtained solutions of short sequences.
Alternatively, we can directly adopt numerical calculations to present the numerical solution.
In the following, we address this issue.

We first study the case of a concatenation of two propagators:
\begin{eqnarray}
U_3({\theta}_1'')=U_2(\bm{\theta}_{2}')U_2(\bm{\theta}'_1)=U_2^{[1]}({\theta}_1^{[1]}),
\end{eqnarray}
where $U_2^{[1]}({\theta}_1^{[1]})$ is used in this subsection for the sake of distinction from the third concatenation in Sec.~SIV.
The equation to be solved reads
\begin{eqnarray}
\partial^{1}_{\delta_\alpha}\!U_{3,eb}=-2ie^{-i 3\alpha } r^* s \alpha\left[e^{i \left(4 \alpha +\theta'_{s,2}\right)}+e^{i \theta' _{s,1}}\right]=0,
\end{eqnarray}
Obviously, the solution of the phase difference is
\begin{eqnarray} \label{s29}
\theta'_{2,1}=\theta'_{s,2}-\theta'_{s,1}=\pi-4\alpha.
\end{eqnarray}
As a result, the sequence is accurate to the second order of the deviation [i.e., $\mathcal{O}(\delta^2_\alpha$)].

By concatenating three propagators, the form of the total propagator is
\begin{eqnarray}
U_3({\theta}_1'')=U_2(\bm{\theta}_{3}')U_2(\bm{\theta}_{2}')U_2(\bm{\theta}'_1),
\end{eqnarray}
where
\begin{eqnarray} \label{9}
\partial^{1}_{\delta_\alpha}\!U_{3,eb}=-2ie^{-i 5\alpha } r^* s \alpha\left[e^{i \left(8 \alpha +\theta'_{s,3}\right)}+e^{i \left(4 \alpha +\theta'_{s,2}\right)}+e^{i \theta' _{s,1}}\right]=0,
\end{eqnarray}
One solution of Eq.~(\ref{9}) can be expressed as
\begin{eqnarray}  \label{10}
\theta'_{21}=2\arctan{\left[\frac{\sqrt{3}-2\sin4\alpha}{2\cos4\alpha-1}\right]}, ~~~~
\theta'_{32}=2\arctan{\left[\frac{\sqrt{3}+2\sin4\alpha}{1-2\cos4\alpha}\right]}.
\end{eqnarray}
Therefore, this sequence is still accurate to the second order of the deviation.

When $N_3>4$, it is difficult to obtain the analytical expression of phase differences by directly solving Eq.~(\ref{s32}).
However, the idea of the concatenation method demonstrated in the main text can be also applied here to obtain the analytical expressions.
For example, there are two ways to obtain the phase differences of the six-propagator sequence. One is directly numerically solving Eq.~(\ref{s32}), and the other is  concatenating Eqs.~(\ref{s29}) and (\ref{10}).
Next, we take $2^M$ propagators $U_2(\bm{\theta}_{n_2}')$ as the example to elaborate the concatenated process.

First of all, we directly regard the propagator $U_2^{[1]}({\theta}_1^{[1]})$ as the new propagator, and then perform two propagators recombination again. In this case, the total propagator becomes
\begin{eqnarray}
U_3({\theta}_1'')=U_2^{[2]}({\theta}_1^{[2]})&=&U_2^{[1]}({\theta}_2^{[1]})U_2^{[1]}({\theta}_1^{[1]}) \\[0.5ex]
&=&U_2(\bm{\theta}_{4}')U_2(\bm{\theta}'_3)U_2(\bm{\theta}_{2}')U_2(\bm{\theta}'_1),
\end{eqnarray}
where the phase difference must satisfy
\begin{eqnarray}
{\theta}_2^{[1]}-{\theta}_1^{[1]}=\pi-8\alpha.
\end{eqnarray}
At this time, the sequence is accurate to the third order of the deviation.

In a similar concatenation way, when the number of pulses is $2^M$, the total propagator reads
\begin{eqnarray}
U_3({\theta}_1'')&=&U_2^{[M]}({\theta}_1^{[M]}) \\[0.5ex]
&=&U_2^{[M-1]}({\theta}_2^{[M-1]}) U_2^{[M-1]}({\theta}_1^{[M-1]}) \\[0.5ex]
&=&\underbrace{U_2^{[M-2]}({\theta}_4^{[M-2]})U_2^{[M-2]}({\theta}_3^{[M-2]})} \underbrace{U_2^{[M-2]}({\theta}_2^{[M-2]})U_2^{[M-2]}({\theta}_1^{[M-2]})}  \\[0.5ex]
&=&\underbrace{\underbrace{U_2^{[M\!-\!3]}(\!{\theta}_8^{[M\!-\!3]}\!)U_2^{[M\!-\!3]}(\!{\theta}_7^{[M\!-\!3]}\!)} \underbrace{U_2^{[M\!-\!3]}(\!{\theta}_6^{[M\!-\!3]}\!)U_2^{[M\!-\!3]}(\!{\theta}_5^{[M\!-\!3]}\!)}} \underbrace{\underbrace{U_2^{[M\!-\!3]}(\!{\theta}_4^{[M\!-\!3]}\!)U_2^{[M\!-\!3]}(\!{\theta}_3^{[M\!-\!3]}\!)} \underbrace{U_2^{[M\!-\!3]}(\!{\theta}_2^{[M\!-\!3]}\!)U_2^{[M\!-\!3]}(\!{\theta}_1^{[M\!-\!3]}\!)}} \nonumber\\  \\[0.5ex]
&=& \cdots    \nonumber\\[0.5ex]
&=&\underbrace{\underbrace{U_2(\bm{\theta}_{2^{M}}')U_2(\bm{\theta}'_{2^{M-1}})} \underbrace{U_2(\bm{\theta}_{2^{M-2}}')U_2(\bm{\theta}'_{2^{M-3}})}}_{\dots} \dots \underbrace{\underbrace{U_2(\bm{\theta}_{4}')U_2(\bm{\theta}'_3)} \underbrace{U_2(\bm{\theta}_{2}')U_2(\bm{\theta}'_1)}}_{\dots}. ~~~~~~~
\end{eqnarray}
The phase differences must satisfy the following relations:
\begin{subequations}  \label{s40}
\begin{eqnarray}
\mathrm{The~first~concatenation}:~&&\theta'_{2n_2,2n_2-1}=\theta_{s,2n_2}-\theta_{s,2n_2-1}=\pi-4\alpha,    \\[0.5ex]
\mathrm{The~second~concatenation}:~&&\theta'_{4n_2-1,4n_2-3}=\theta_{s,4n_2-1}-\theta_{s,4n_2-3}=\pi-8\alpha,  \\[0.5ex]
\mathrm{The~third~concatenation}:~&&\theta'_{8n_2-3,8n_2-7}=\theta_{s,8n_2-3}-\theta_{s,8n_2-7}=\pi-16\alpha,  \\[0.5ex]
\dots ~~~~~~~~~~~~~~~~~~~&& \nonumber\\[0.5ex]
\mathrm{The~\emph{M}^{th}~concatenation}:~&&\theta'_{2^Mn_2-2^M+1,2^Mn_2-2^{M-1}+1}= \theta_{s,2^Mn_2-2^M+1}-\theta_{s,2^Mn_2-2^{M-1}+1}=\pi-2^{M+1}\alpha,~~~~~~
\end{eqnarray}
\end{subequations}
where the sequence has a deviation compensation up to the order $O(\delta_\alpha^{M+1})$.
Notice that the first phase $\theta_{1}$ of the first SP pulse pair is arbitrary, which can be modulated for other hierarchical concatenations (e.g., the third concatenation in the main text).

Figure~\ref{ueb} presents the amplitude of the element $|U_{N_2,eb}|$ against the deviation $\delta_{\alpha}$ for different $N_2$, where the case from $N_2=4$ to $N_2=8$ is derived by numerical calculations.
As shown by the curves in Fig.~\ref{ueb}(a), the robust performance of transitionless quantum driving becomes superior overall with the pulse number increasing.
On the other hand,
we can also find that the $N_2=2$ sequence has a better robustness than the $N_2=3$ sequence.
This is because both of their Taylor series are only ensured to be accurate to the second order of $\delta_{\alpha}$,
while the amplitude of $\delta_{\alpha}^2$ in the former is smaller than the latter.
Therefore, the paired pulses perform more excellent in the second concatenation, and a similar result is also obtained by a comparison between $N_2=4$ and $N_2=5,6$.
In Fig.~\ref{ueb}(b),
we compare the robustness performance achieved by the numerical method with the concatenated method.
Generally speaking,
both methods can obtain a similar robust performance of transitionless quantum driving.
While it is more convenient and efficient to acquire all phases by merely matching the phase differences from Eq.~(\ref{s40}) instead of finding the solution via lots of tedious numerical calculations.

\begin{figure}[htbp]
	\centering
	\includegraphics[scale=0.53]{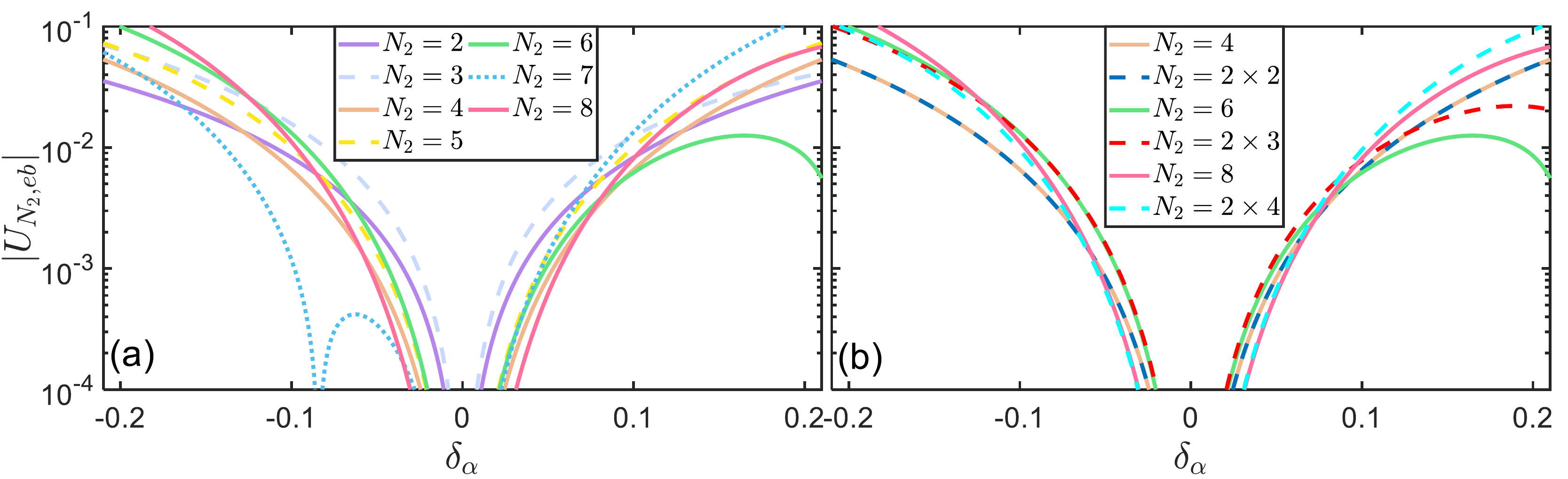}
	\caption{The amplitude of the element $|U_{N_2,eb}|$ vs the deviation $\delta_{\alpha}$ by different $N_2$ sequences.
We set $r=0.1$ and $\alpha=1.029$, and all phase differences is presented in Table ~\ref{tabueb}.}\label{ueb}
\end{figure}

\begin{table}[htbp]
\centering
\caption{Phase differences for the second concatenation of SP pulse pairs by the numerical method.}\label{tabueb}
\setlength\tabcolsep{17pt}
 \begin{tabular}{cccccccccccccccc}
 \hline
 \hline
 \multirow{2}{*}{$N_2$}&\multirow{2}{*}{$\theta'_{21}$}&\multirow{2}{*}{$\theta'_{32}$}& \multirow{2}{*}{$\theta'_{43}$}&\multirow{2}{*}{$\theta'_{54}$}&\multirow{2}{*}{$\theta'_{65}$} &\multirow{2}{*}{$\theta'_{76}$}&\multirow{2}{*}{$\theta'_{87}$}\\\\
 \hline
 \multirow{2}{*}{4}& \multirow{2}{*}{ $-$0.9750}& \multirow{2}{*}{2.1678}& \multirow{2}{*}{$-$0.9744}& \multirow{2}{*}{--}& \multirow{2}{*}{--}& \multirow{2}{*}{--}& \multirow{2}{*}{--} \\\\
 \multirow{2}{*}{5}& \multirow{2}{*}{4.8034}& \multirow{2}{*}{$-$2.2925}& \multirow{2}{*}{0.3437}& \multirow{2}{*}{$-$0.4690}& \multirow{2}{*}{--}& \multirow{2}{*}{--}& \multirow{2}{*}{--} \\\\
 \multirow{2}{*}{6}& \multirow{2}{*}{5.0645}& \multirow{2}{*}{$-$4.4539}& \multirow{2}{*}{$-$1.3636}& \multirow{2}{*}{1.8294}& \multirow{2}{*}{$-$1.2188}& \multirow{2}{*}{--}& \multirow{2}{*}{--} \\\\
 \multirow{2}{*}{7}& \multirow{2}{*}{$-$0.7739}& \multirow{2}{*}{$-$3.8410}& \multirow{2}{*}{0.2731}& \multirow{2}{*}{0.2287}& \multirow{2}{*}{2.6760}& \multirow{2}{*}{$-$0.8487}& \multirow{2}{*}{--} \\\\
 \multirow{2}{*}{8}& \multirow{2}{*}{$-$1.1366}& \multirow{2}{*}{1.8387}& \multirow{2}{*}{$-$2.0145}& \multirow{2}{*}{$-$3.3636}& \multirow{2}{*}{4.2688}& \multirow{2}{*}{$-$4.4445}& \multirow{2}{*}{5.1465} \\\\
\hline
\hline
 \end{tabular}
\end{table}

%\clearpage
\section{High precision quantum operations}

After the second concatenation of SP pulse pairs, the nonadiabatic transition induced by
various errors is dramatically suppressed, and thus the propagator $U_3({\theta}''_1)$ can be approximately written in a diagonalized form,
\begin{eqnarray} \label{s44}
U_3({\theta}''_1)\approx|d(t)\rangle \langle d(t)|+\exp[{i2\beta(1+\delta_{\beta})}]|b(t)\rangle \langle b(t)|+\exp[{-i2\beta'(1+\delta_{\beta'})}]|e\rangle \langle e|.
\end{eqnarray}
Actually, the deviation $\delta_\beta$ in Eq.~(\ref{s44}) would decline the precision of quantum operations.
To show it more clearly, we transform the propagator $U_3({\theta}''_1)$ back to the original representation spanned by $\{|g\rangle, |f\rangle, |e\rangle\}$.
In the basis $\{|g\rangle, |f\rangle\}$, the form of the propagator $U_3({\theta}''_1)$ is given by
\begin{eqnarray}\label{errant45}
U_3({\theta}''_1)=\exp\left[{-\frac{i}{2}\beta(1+\delta_{\beta})\bm{n}\cdot\bm{\sigma}}\right],
\end{eqnarray}
where $\bm{n}=(-\sin\phi(t)\cos\theta_{sp}, \sin\phi(t)\sin\theta_{sp}, \cos\phi(t))$.
It is easily observed from Eq.~(\ref{errant45}) that the rotation angle would produce errors in the Bloch sphere due to the presence of the deviation $\delta_\beta$.

To further restrain the influence of the deviation $\delta_\beta$ on the precision of quantum operations, we need to a third
concatenation of SP pulse pairs.
That is, we concatenate ${N_3}$ propagators $U_3({\theta}_{n_3}'')$ to form a new one,
\begin{eqnarray}\label{pro4}
U_4({\theta}_1''')=U_3({\theta}_{N_3}'')\cdots U_3({\theta}''_1)=\left[
\begin{array}{ccc}
                 {U}_{4,gg} &{U}_{4,gf}          &  0  \\[1ex]
                 {U}_{4,fg}&{U}_{4,ff}&0 \\[1ex]
                 0                    &0&1
\end{array}
\right],
\end{eqnarray}
where ${\theta}''_{n_3}$ is the adjustable phases in this situation, $n_3=1, \dots, N_3$.
{Next,
we design the phases ${\theta}''_{n_3}$ by using two different ways.
The first one is to compensate for the deviation only in the population of the target state,
called as the population-compensation (PC) sequence.
The second one is to compensate for the deviation in the both population and phase of the target state,
called as fidelity-compensation (FC) sequence hereafter.}

The construction of the PC sequence is as follows.
Taking similar steps in Sec.~\ref{siii}, we modulate the phases ${\theta}_{n_3}''$ to nullify the coefficients $\partial^{k}_{\delta_\beta}\!|U_{4,fg}|^2$ in  a increasing order as many as possible, where $\partial^{k}_{\delta_\beta}\!|U_{4,fg}|^2$ represents the $k^\mathrm{th}$ derivative of $|U_{4,fg}|^2$ with respect to $\delta_\beta$, $k=1,2,\dots$
It is worth mentioning that the zero-order coefficient $\partial^{0}_{\delta_\beta}\!|U_{4,fg}|^2$ is used to control the population of the state $|f\rangle$.
In the following, we first take the case $N_3=3$ as the example to demonstrate the design of the phases ${\theta}''_{n_3}$.
The form of the total propagator reads
\begin{eqnarray}
U_4({\theta}_1''')=U_3(\theta_3'')U_3(\theta_2'')U_3(\theta_1''),
\end{eqnarray}
and the equations to be solved are
\begin{subequations}\label{s49}
\begin{eqnarray}
\partial^{0}_{\delta_\beta}\!|U_{4,fg}|^2&=&\frac{1}{4}\left[2+\cos(\theta_{21}''+\theta_{32}'')- \cos(\theta_{21}''-\theta_{32}'')\right]=P_f,\\
\partial^{1}_{\delta_\beta}\!|U_{4,fg}|^2&=&-\frac{\pi}{4}\left[2 \cos\theta_{32}'' \cos ^2\frac{\theta_{21}''}{2}+\cos \theta_{21}''\right]=0,
\end{eqnarray}
\end{subequations}
where $P_f$ represents the population of the state $|f\rangle$.
It readily obtains the solution of Eq.~(\ref{s49}), which are
\begin{subequations}
\begin{eqnarray}
\theta_{21}''&=&2\arctan \pm\left(\sqrt{1-P_f^2}\pm\sqrt{2P_f-P_f^2}\right), \\
\theta_{32}''&=&2\arctan \pm\left(\sqrt{1-P_f^2}\mp\sqrt{2P_f-P_f^2}\right).
\end{eqnarray}
\end{subequations}
When ${N_3}$ is large, it is convenient to adopts numerical methods, and a group of numerical solutions for different $N_3$ are presented in Table~\ref{tab3}, while the corresponding excitation profiles of the state $|f\rangle$ are plotted in Fig.~\ref{deltabeta}(a).
Obviously,
the excitation profile for the single propagator $U_3({\theta}_{1}'')$ cannot keep the target value in the presence of the deviation $\delta_{\beta}$,
while the excitation profile for the case $N_3=3$ shows a primary flat platform around $\delta_{\beta}=0$.
To achieve better robustness,
we can continue to concatenate more propagators $U_3({\theta}_{n_3}'')$ to generate a longer sequence.
Of course, the long sequence can also be concatenated by short sequences.
In this way,
more free phases ${\theta}''_{n_3}$ are introduced to nullify more high-order coefficients $\partial^{k}_{\delta_\beta}\!|U_{4,fg}|^2$.
As expected,
the flat central platform of the excitation profile around $\delta_{\beta}=0$ becomes wider with the increase of $N_3$; see the solid yellow and violet curves in Fig.~\ref{deltabeta}(a).

\begin{figure}[htbp]
\centering
\includegraphics[scale=0.548]{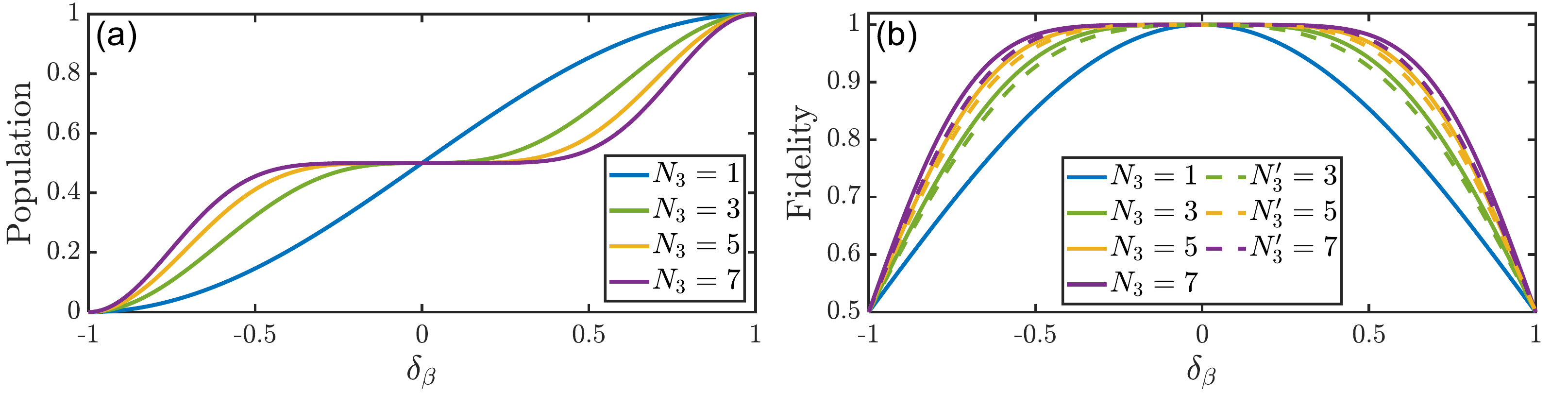}
\caption{{(a) Population $P_f$ vs the deviation $\delta{\beta}$.
The target population of the state $|f\rangle$ is $P_f$=0.5, $\phi(t)=\pi/4$, and other parameters are given in Table~\ref{tab3}.
(b) Fidelity $F$ vs the deviation $\delta{\beta}$. The parameters are given in Table~\ref{tab4}, and the target state is the maximum superposition state $|\psi_T\rangle=1/\sqrt{2}(|g\rangle+|f\rangle)$. Dashed (solid) curves represent the excitation profiles designed by fidelity (population) formulas.
}}\label{deltabeta}
\end{figure}

\begin{table}[htbp]
\centering
\caption{Phase differences  for the PC sequence in the third concatenation.}\label{tab3}
\setlength\tabcolsep{15pt}
 \begin{tabular}{cccccccccccccccc}
 \hline
 \hline
 \multirow{2}{*}{$N_3$}&\multirow{2}{*}{$\theta''_{21}$}&\multirow{2}{*}{$\theta''_{32}$}& \multirow{2}{*}{$\theta''_{43}$}&\multirow{2}{*}{$\theta''_{54}$}&\multirow{2}{*}{$\theta''_{65}$}&\multirow{2}{*}{$\theta''_{76}$}\\\\
 \hline
 \multirow{2}{*}{3}& \multirow{2}{*}{2.0944}& \multirow{2}{*}{0}& \multirow{2}{*}{--} & \multirow{2}{*}{--} & \multirow{2}{*}{--}& \multirow{2}{*}{--}\\\\
% \multirow{2}{*}{4}& \multirow{2}{*}{2.3562}& \multirow{2}{*}{0}& \multirow{2}{*}{0.7854}& \multirow{2}{*}{--} \\\\
 \multirow{2}{*}{5}& \multirow{2}{*}{2.5133}& \multirow{2}{*}{0}& \multirow{2}{*}{1.2566}& \multirow{2}{*}{0}& \multirow{2}{*}{--}& \multirow{2}{*}{--} \\\\
 \multirow{2}{*}{7}& \multirow{2}{*}{2.6928}& \multirow{2}{*}{0}& \multirow{2}{*}{1.7952}& \multirow{2}{*}{0}& \multirow{2}{*}{0.8976}& \multirow{2}{*}{0} \\\\
\hline
\hline
 \end{tabular}
\end{table}

{

To robustly obtain an arbitrary superposition state $|\psi_T\rangle=\cos{\alpha}|g\rangle+\sin{\alpha}e^{i\chi}|f\rangle$,
it is not enough to achieve high-precision population transfer because different relative phases $\chi$ correspond to distinct states even if they have the same $\alpha$.
Therefore,
if the objective is to implement a superposition state carrying an specific relative phase,
it is inadequate to merely exploit the Taylor expansion of the amplitude $|U_{4,fg}|^2$.
Next,
we describe how to design such pulse sequence that takes the relative phase into account by using
the fidelity $F$ instead, whose definition is given by
\begin{eqnarray}  \label{sfidelity}
F=|\langle\psi_T|\Psi(N'_3T)\rangle|^2,
\end{eqnarray}
where $|\Psi(N'_3T)\rangle$ represents the final state after concatenating $N'_3$ pulses.
Here, to make it distinct to the PC sequence, we replace the pulse number $N_3$ by $N'_3$.

Again, starting from the propagator~(\ref{pro4}),
when the initial state of the system is $|g\rangle$,
the final state becomes
\begin{eqnarray}
|\psi_T\rangle=U_{4,gg}|g\rangle+U_{4,fg}|f\rangle.
\end{eqnarray}
As a result,
the expression of the fidelity $F$ in Eq.~(\ref{sfidelity}) reads
\begin{eqnarray}
F=|U_{4,gg}|^2\cos^2{\alpha}+|U_{4,fg}|^2\sin^2{\alpha}+\dfrac{1}{2}\sin{2\alpha}\left(U_{4,gg}U^*_{4,fg}e^{i\chi}+U_{4,fg}U^*_{4,gg}e^{-i\chi}\right).
\end{eqnarray}
In the presence of the deviation,
the Taylor expansion of the fidelity can be written as
\begin{eqnarray} \label{sf50}
F&=&f_0+f_1\delta_\beta+f_2\delta_\beta^2+\cdots,
\end{eqnarray}
where $f_{k'}=(k'!)^{-1}\partial^{k'}_{\delta_\beta}F|_{\delta_\beta=0}$ is the coefficient of the $k'^{\mathrm{th}}$-order term.
At this time,
in order to obtain a superposition state with high fidelity while maintaining robust against the deviation $\delta_\beta$,
it is necessary to vanish as more low-order terms in Eq.~(\ref{sf50}) as possible.

We begin with exemplifying this pulse design by using $N'_3=3$, where only the zero-order and first-order terms need to be concerned. The expressions of these two coefficients are given by
\begin{subequations}
\begin{eqnarray}
f_0&=&\frac{1}{16}\Big\{8+\cos{(\chi\!+\!\theta_1''\!-\!2\alpha)}-\cos{(\chi\!+\!\theta_1''\!+\!2\alpha)}- \cos{(\chi\!+\!\theta_1''\!-\!2\alpha\!+\!2\theta_{21}'')}+8\cos{2\alpha}\sin{\theta_{21}''}\sin{\theta_{32}''}\nonumber\\
&&-\cos{(\chi\!+\!\theta_1''\!+\!2\alpha\!+\!2\theta_{21}'')}-4\sin{2\alpha}\big[\cos{(\chi\!+\!\theta_1'' \!+\!\theta_{21}''\!+\!2\theta_{32}'')}+2\cos{\theta_{21}''}\sin{(\chi\!+\!\theta_1''\!+\!\theta_{21}''\! +\!\theta_{32}'')}\big]\Big\},\label{f0}\\[1ex]
f_1&=&\frac{\pi}{16}\big\{2\sin{2\alpha}\left[\sin{(\chi\!+\!\theta_1''\!+\!\theta_{21}'')} \!-\!\sin{(\chi\!+\!\theta_1''\!+\!2\theta_{32}'')}\!+\!\sin{(\chi\!+\!\theta_1''\!+\!\theta_{21}''\! +\!2\theta_{32}'')}\right]\!+\!2\cos{(2\alpha+\theta_{21}'')}\nonumber\\
&&+2\cos{(2\alpha\!-\!\theta_{21}'')}+\cos{(\chi+\theta_1''\!+\!2\alpha)}-\cos{(\chi\!+\!\theta_1''\!-\!2\alpha)} +8\cos{2\alpha}\cos^2{\frac{\theta_{21}''}{2}}\cos{\theta_{32}''}\big\},\label{f1}
\end{eqnarray}
\end{subequations}
where the phase differences are $\theta_{21}''=\theta''_2-\theta''_1$ and $\theta_{32}''=\theta''_3-\theta''_2$.
Hence,
the equations to be solved are
\begin{subequations} \label{s52ab}
\begin{numcases}{}
f_0=1,\label{f0=1}\\[1ex]
f_1=0,\label{f1=0}
\end{numcases}
\end{subequations}
where Eq.~(\ref{f0=1}) ensures to perfectly obtain  the target state in the absence of the deviation $\delta_\beta$,
and Eq.~(\ref{f1=0}) means the nullification of the first-order term.
Generally,
obtaining analytical solutions for Eqs.~(\ref{s52ab}) is quite difficult except for some special circumstances.
For instance,
when $\alpha=\pi/4$,
i.e.,
the target state being the maximum superposition state, Eqs.~(\ref{s52ab}) can be reduced into
\begin{subequations} \label{s53ab}
\begin{numcases}{}
\frac{1}{8}\Big[4\!+\!\sin{\zeta}\!-\!\sin{(2\theta_{21}''\!+\!\zeta)} -2\sin(\theta_{21}'')\cos(\theta_{21}''+2\theta_{32}''+\zeta) -4\cos(\theta_{21}'')\sin(\theta_{21}''+\theta_{32}''+\zeta)\Big]=1,\label{f0==1}\\[1ex]
\frac{\pi}{2}\cos{\frac{\theta_{21}''}{2}}\sin{\theta_{32}''} \cos{(\frac{\theta_{21}''}{2}+\theta_{32}''+\zeta)}=0,\label{f1==0}
\end{numcases}
\end{subequations}
where $\zeta=\theta_1''+\chi$.
Obviously, it is possible to attain the analytical solutions of Eq.~(\ref{s53ab}), which are given in the Table~\ref{tabsol}.
It is worth mentioning that although there may be many solutions to the phases, we usually select a set of solutions that have less effect on the deviation; this can be done by inputting  all the solutions into the second-order coefficient $f_2$ and find out  the ones that result in the smallest value of $f_2$.

\begin{table}[htbp]
{
\centering
\caption{Analytical solutions of the phases for $\alpha=\pi/4$ in the FC sequence ($m$ is an arbitrary integer).}\label{tabsol}
\setlength\tabcolsep{26pt}
 \begin{tabular}{cccccccccccccccc}
 \hline
 \hline
 \multirow{2}{*}{$\theta''_{21}$}&\multirow{2}{*}{$\theta''_{32}$}& \multirow{2}{*}{$\theta_1''$}\\\\
 \hline\\[-1ex]
  \multirow{2}{*}{$m\pi$}& \multirow{2}{*}{$\dfrac{3\pi}{2}-\theta_1''-\chi+2m\pi$}& \multirow{2}{*}{arbitrary}\\\\[2.5ex]
  \multirow{2}{*}{$\dfrac{3\pi}{4}-\dfrac{\theta_1''+\chi}{2}+m\pi$}& \multirow{2}{*}{$2m\pi$}& \multirow{2}{*}{arbitrary}\\\\[1ex]
  \multirow{2}{*}{arbitrary}&\multirow{2}{*}{$\pi+2m\pi$}& \multirow{2}{*}{$\dfrac{\pi}{2}-\chi+2m\pi$}\\\\[2.5ex]
  \multirow{2}{*}{$\pm\dfrac{4\pi}{3}+4m\pi$}&\multirow{2}{*}{$2m\pi$}& \multirow{2}{*}{$\dfrac{3\pi}{2}\mp\dfrac{2\pi}{3}-\chi+2m\pi$}\\\\[1.5ex]
      \multirow{2}{*}{$\pm\dfrac{2\pi}{3}+4m\pi$}&\multirow{2}{*}{$2m\pi$}& \multirow{2}{*}{$\dfrac{\pi}{2}\mp\dfrac{\pi}{3}-\chi+2m\pi$}\\\\[1.5ex]
\hline
\hline
 \end{tabular}
 }
\end{table}

Based on the complexity of analytical solutions, we tend to numerical solutions for the FC sequence,
and the phase differences of $N'_3=5$ and $N'_3=7$ are presented in Table~\ref{tab4}.
Meanwhile,
we plot in Fig.~\ref{deltabeta}(b) the fidelity $F$ as a function of the deviation $\delta_{\beta}$ for the FC sequence with different pulse numbers.
The results show that,
the FC sequence can effectively suppress the phase sensitivity of the target state, and possess remarkable deviation compensation capability that the single pulse sequence lacks of. In particular, the robustness becomes much better as the pulse number increases.
As a contrast,
the fidelity for the PC sequence is also plotted in Fig.~\ref{deltabeta}(b).
In terms of robustness against the deviation $\delta_{\beta}$, the PC sequence outperforms the FC sequence at the same number of pulses.
The reason is that the FC sequence requires to nullify the deviation in the both population and phase for achieving desired superposition states,
implying that more phases and thus a longer sequence are involved to nullify the same order of deviation compensation.

\begin{table}[htbp]
{
\centering
\caption{Phase differences for the FC sequence in the third concatenation.}\label{tab4}
\setlength\tabcolsep{15pt}
 \begin{tabular}{cccccccccccccccc}
 \hline
 \hline
 \multirow{2}{*}{$N'_3$}&\multirow{2}{*}{$\theta''_{21}$}&\multirow{2}{*}{$\theta''_{32}$}& \multirow{2}{*}{$\theta''_{43}$}&\multirow{2}{*}{$\theta''_{54}$}&\multirow{2}{*}{$\theta''_{65}$}&\multirow{2}{*}{$\theta''_{76}$}\\\\
 \hline
 \multirow{2}{*}{3}& \multirow{2}{*}{2.3562}& \multirow{2}{*}{0}& \multirow{2}{*}{--} & \multirow{2}{*}{--} & \multirow{2}{*}{--}& \multirow{2}{*}{--}\\\\
% \multirow{2}{*}{4}& \multirow{2}{*}{2.3562}& \multirow{2}{*}{0}& \multirow{2}{*}{0.7854}& \multirow{2}{*}{--} \\\\
 \multirow{2}{*}{5}& \multirow{2}{*}{1.5708}& \multirow{2}{*}{2.3562}& \multirow{2}{*}{-1.5708}& \multirow{2}{*}{-1.5708}& \multirow{2}{*}{--}& \multirow{2}{*}{--} \\\\
 \multirow{2}{*}{7}& \multirow{2}{*}{3.1416}& \multirow{2}{*}{-2.3886}& \multirow{2}{*}{0.0097}& \multirow{2}{*}{2.6366}& \multirow{2}{*}{-0.0096}& \multirow{2}{*}{-1.8235} \\\\
\hline
\hline
 \end{tabular}
 }
\end{table}
}

\section{Implementation of robust transitionless quantum driving by different pulse shapes}\label{appdb}

As we have demonstrated in the main text, the current approach is suitable for arbitrary pulse shapes. In the main text, we employ the Gaussian shape to implement a specific rotation operation, where the relevant parameters can be found in Table~\ref{gauss3}. Here, we illustrate other typical and regular pulse shapes, such as the sinusoidal pulse, the sawtooth pulse, the triangle pulse, and the trapezoidal pulse.
In Table~\ref{t1},
we give out the function expressions for the Stokes pulse, where the parameter $\tau$ denotes the time reaching the peak of the pulse.
In the trapezoidal pulse,
$\tau_1$ and $\tau_2$ are the start and final moment of the peak value.
For simplicity,
we set the detuning as constant.
Certainly,
it is feasible to simultaneously set the coupling strength and detuning as time-dependent functions as well.

\begin{table}[htbp]
\centering
\caption{The relevant parameters for the $\mathcal{R}_{\frac{\pi}{4}}(N_1\times N_2,N_3)$ sequence with the Gaussian shape (in units of $1/\Omega_0$), where $\theta_{p,1}=\theta_{s,1}$.}\label{gauss3}
\setlength\tabcolsep{12pt}
 \begin{tabular}{ccccccccccccccccc}
 \hline
 \hline
 \multirow{2}{*}{$\mathcal{R}_{\frac{\pi}{4}}(N_1\times N_2,N_3)$}&\multirow{2}{*}{$A$}&\multirow{2}{*}{$\Delta$}&\multirow{2}{*}{$\alpha$}&\multirow{2}{*}{$\theta_{s,1}$}\\\\
 \hline
\multirow{2}{*}{$\mathcal{R}_{\frac{\pi}{4}}(2,1)$}&\multirow{2}{*}{1.099}&\multirow{2}{*}{0.6574}&\multirow{2}{*}{1.1868}&\multirow{2}{*}{1.5708}\\\\
\multirow{2}{*}{$\mathcal{R}_{\frac{\pi}{4}}(4,3)$}&\multirow{2}{*}{2.381}&\multirow{2}{*}{0.2802}&\multirow{2}{*}{0.4479}&\multirow{2}{*}{0.5237}\\\\
\multirow{2}{*}{$\mathcal{R}_{\frac{\pi}{4}}(4,5)$}&\multirow{2}{*}{2.381}&\multirow{2}{*}{0.2802}&\multirow{2}{*}{0.4479}&\multirow{2}{*}{$-$0.9425}\\\\
\multirow{2}{*}{$\mathcal{R}_{\frac{\pi}{4}}(8,1)$}& \multirow{2}{*}{1.1299}&\multirow{2}{*}{0.1640}&\multirow{2}{*}{0.2957}&\multirow{2}{*}{1.5708}\\\\
\multirow{2}{*}{$\mathcal{R}_{\frac{\pi}{4}}(8,5)$}& \multirow{2}{*}{1.1299}&\multirow{2}{*}{0.1640}&\multirow{2}{*}{0.2957}&\multirow{2}{*}{$-$0.9425}\\\\
\hline
\hline
 \end{tabular}
\end{table}

\begin{table}[htbp]
	\centering
	\caption{The expressions of some common waveforms.}
	\label{t1}
	\begin{tabular}{cl}
		\hline
		\hline\\[-2ex]
         {Waveforms}  & ~~~~~~~~~{Expressions}  \\[.5ex]
        \hline\\[-1ex]
        Gaussian  &~~~~~~~~ $\Omega_{s}(t)=A\Omega_0e^{-(t-\tau)^2/T^2}$  \\[2ex]
		
        Sinusoidal  &~~~~~~~~ $\Omega_{s}(t)=A\Omega_0|\sin(\pi t/T+\tau)|$  \\[2ex]
		        \specialrule{0em}{3pt}{3pt}
        Sawtooth  &~~~~~~~~ $\Omega_{s}(t)\!=\!\displaystyle\frac{A\Omega_0}{T}t,  ~~~~~~~~~~~~~~~~~~~~~~~0\leq t\leq T$  \\[2ex]
        \specialrule{0em}{3pt}{3pt}
        Triangle  &~~~~~~~~ $\Omega_{s}(t)\!=\!\left\{
                                \begin{aligned}
                                     &\displaystyle\frac{A\Omega_0}{\tau}t,  ~~~~~~~~~~~~~~~~~~~~0\leq t\leq\tau \\[1ex]
                                     &\displaystyle\frac{A\Omega_0}{\tau-T}(t-T),  ~~~~~~~~~~\tau< t\leq T \\[1ex]
                                \end{aligned}
                                \right.$  \\[2ex]

        \specialrule{0em}{3pt}{3pt}
        Trapezoidal  &~~~~~~~~ $\Omega_{s}(t)\!=\!\left\{
                               \begin{aligned}
                               &\displaystyle\frac{A\Omega_0}{\tau_1}t,  ~~~~~~~~~~~~~~~~~~~~0\leq t\leq\tau_1 \\[1ex]
                               &A\Omega_0,  ~~~~~~~~~~~~~~~~~~~~~\tau_1< t\leq\tau_2 \\[1ex]
                               &\displaystyle\frac{A\Omega_0}{\tau_1+\tau_2-T}(t-T),  ~~\tau_2< t\leq T \\[1ex]
                               \end{aligned}
                               \right.$  \\[3ex]
        %Sawtooth  &~~ $\Delta(t)\!=\!\displaystyle\frac{A\Delta_0}{T}t,  ~~~~~~~~~~~~~~~~~~~~~~~~0\leq t\leq T$  \\[2ex]
		\\[-1ex]\hline
		\hline
	\end{tabular}
\end{table}

We first simulate perfect transitionless quantum driving by different pulse shapes without systematic errors.
Specifically,  we plot in Fig.~\ref{pulse_shape} the population evolution of three states, and the corresponding waveforms  of the Stokes pulse, where the target state is $|\Psi_T\rangle=1/\sqrt{2}(|g\rangle+|f\rangle)$.
At the evolution time $t=2T$,
regardless of the waveform of the SP pulse pair,
the maximum superposition of the state $|g\rangle$ and $|f\rangle$ can be successfully achieved.
This result gives the evidence that the leakage to the state $|e\rangle$ has been completely forbidden by matching the phase difference $\theta_{21}=\pi-2\alpha$.

\begin{figure*}[htbp]
	\centering
	\includegraphics[scale=0.5]{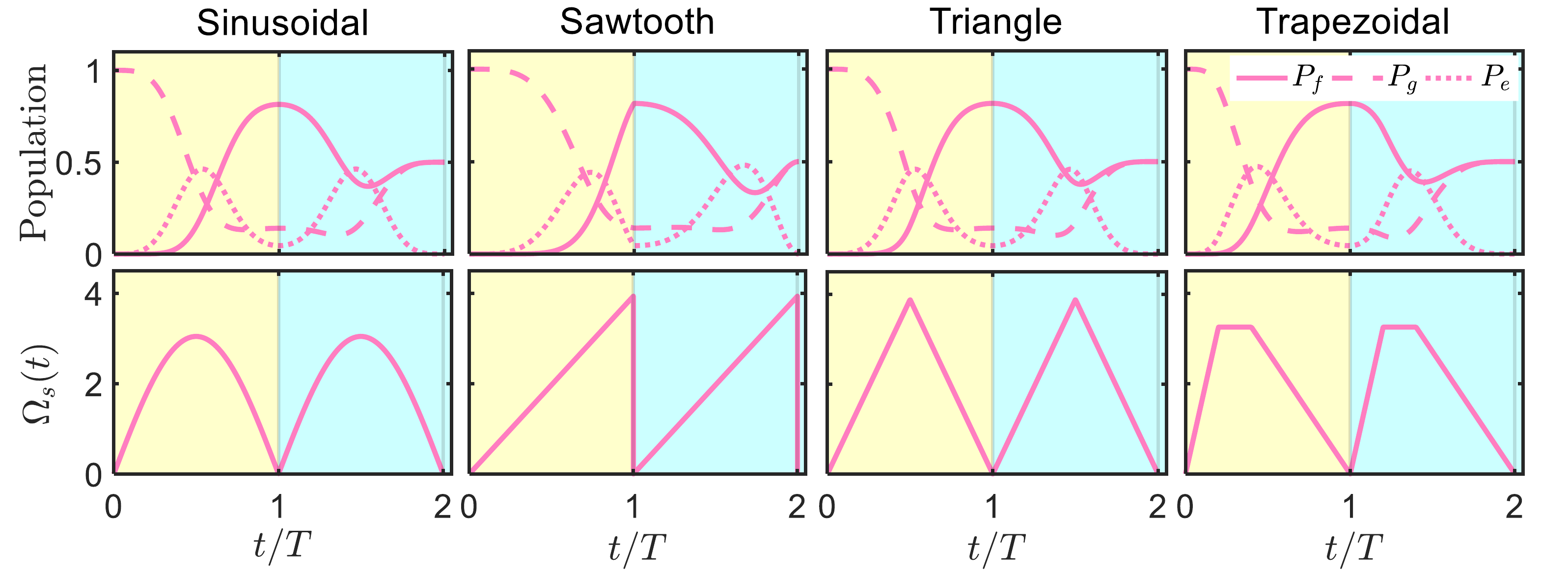}
	\caption{
Population evolution (top panels) of three states and corresponding waveform (bottom panels) of the Stokes pulse by two SP pulse pairs. The target state is $|\Psi_T\rangle=1/\sqrt{2}(|g\rangle+|f\rangle)$, and the relevant parameters of different shapes are given in Table~\ref{TABIV}. Definitely,  all shapes can be used for perfect transitionless quantum driving. }\label{pulse_shape}
\end{figure*}

\begin{table}[htbp]
\centering
\caption{The parameter for different pulse shapes (in units of $1/\Omega_0$).}\label{TABIV}
\setlength\tabcolsep{27pt}
 \begin{tabular}{cccccccccccccccc}
 \hline
 \hline
 \multirow{2}{*}{Waveforms}&\multirow{2}{*}{$A$}&\multirow{2}{*}{$\Delta$}&\multirow{2}{*}{$T$}&\multirow{2}{*}{$\tau$}&\multirow{2}{*}{$\tau_2$}\\\\
 \hline
\multirow{2}{*}{Sinusoidal}&\multirow{2}{*}{3.044}&\multirow{2}{*}{1.98}&\multirow{2}{*}{1}&\multirow{2}{*}{0}&\multirow{2}{*}{--}\\\\
\multirow{2}{*}{Sawtooth}&\multirow{2}{*}{3.935}&\multirow{2}{*}{1.865}&\multirow{2}{*}{1}&\multirow{2}{*}{--}&\multirow{2}{*}{--}\\\\
\multirow{2}{*}{Triangle}&\multirow{2}{*}{3.884}&\multirow{2}{*}{2.18}&\multirow{2}{*}{1}&\multirow{2}{*}{0.5}&\multirow{2}{*}{--}\\\\
\multirow{2}{*}{Trapezoidal}&\multirow{2}{*}{3.249}&\multirow{2}{*}{2.04}&\multirow{2}{*}{1}&\multirow{2}{*}{0.2}&\multirow{2}{*}{0.2}\\\\
\hline
\hline
 \end{tabular}
\end{table}

When the system exhibits the errors,
we take the $\mathcal{R}_{\frac{\pi}{4}}(8,5)$ sequence to demonstrate the feasibility of the robustness with respect to the errors by different shapes in Fig.~\ref{pulse_shape2},
where $\mathcal{R}_{\frac{\pi}{4}}(8,5)$ represents the rotation around the axis $\bm{n}$ with the angle ${{\pi}/{4}}$.
For different pulse shapes,
it is available to achieve the high-fidelity quantum operation with favorable robustness against systematic errors,
as shown by the red regions on the top panel in Fig.~\ref{pulse_shape2}.

\begin{figure*}[htbp]
	\centering
	\includegraphics[scale=0.6]{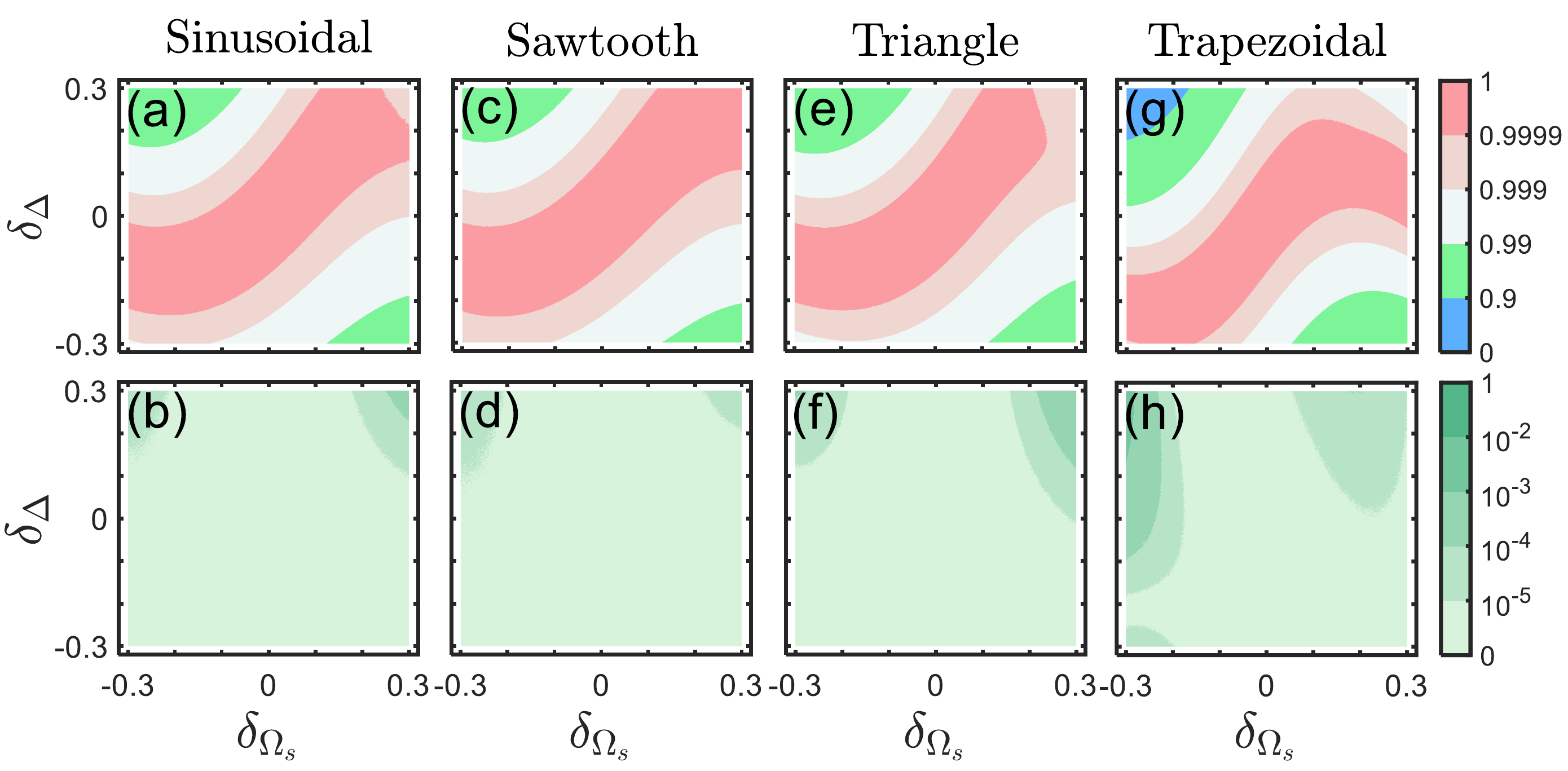}\label{pulse_shape2}
	\caption{
Fidelity $F$ of the target state $|\Psi_T\rangle=1/\sqrt{2}(|g\rangle+|f\rangle)$ (top panels) and population $P_e$ of the state $|e\rangle$ (bottom panels) vs pulse amplitude error $\delta_{\Omega_s}$ and detuning error $\delta_\Delta$ by the $\mathcal{R}_\frac{\pi}{4}(8,5)$ sequence with (a--b) the sinusoidal, (c--d) the sawtooth, (e--f) the triangle, and (g--h) the trapezoidal pulses. The parameters of different shapes are given in Table~\ref{TABS4}.}
\end{figure*}

\begin{table}[htbp]
\centering
\caption{The parameters for different pulse shapes (in units of $1/\Omega_0$).}\label{TABS4}
\setlength\tabcolsep{20pt}
 \begin{tabular}{cccccccccccccccc}
 \hline
 \hline
 \multirow{2}{*}{Waveforms}&\multirow{2}{*}{$A$}&\multirow{2}{*}{$\Delta$}&\multirow{2}{*}{$\tau$}&\multirow{2}{*}{$\tau_2$}&\multirow{2}{*}{$\alpha$}&\multirow{2}{*}{$\theta_{p,1}$}\\\\
 \hline
\multirow{2}{*}{Sinusoidal}&\multirow{2}{*}{3.1445}&\multirow{2}{*}{0.496}&\multirow{2}{*}{3}&\multirow{2}{*}{--}&\multirow{2}{*}{0.0518}&\multirow{2}{*}{$-$0.9425}\\\\
\multirow{2}{*}{Sawtooth}&\multirow{2}{*}{4.0070}&\multirow{2}{*}{0.46}&\multirow{2}{*}{3}&\multirow{2}{*}{--}&\multirow{2}{*}{0.0337}&\multirow{2}{*}{$-$0.9416}\\\\
\multirow{2}{*}{Triangle}&\multirow{2}{*}{4.0040}&\multirow{2}{*}{0.546}&\multirow{2}{*}{0.5}&\multirow{2}{*}{--}&\multirow{2}{*}{0.0766}&\multirow{2}{*}{$-$0.9416}\\\\
\multirow{2}{*}{Trapezoidal}&\multirow{2}{*}{4.0535}&\multirow{2}{*}{0.6615}&\multirow{2}{*}{0.2}&\multirow{2}{*}{0.2}&\multirow{2}{*}{0.1343}&\multirow{2}{*}{$-$0.9416}\\\\
\hline
\hline
 \end{tabular}
\end{table}

\section{Gauge invariance for all Hamiltonians}\label{appdb}

In the main text, we briefly demonstrate the property of the Hamiltonian with near-neighbor interactions is trivial when the constant phases is arbitrary.
In fact, this can be suitable for \emph{all} Hamiltonians, up to some constraints on the phases. Next, we explain this issue.

The general form of the time-dependent Hamiltonian reads
\begin{eqnarray}\label{s49a}
H(t)&=&\sum_{j=1}^{J}\lambda_{j,j}(t)|j\rangle\langle j|+\sum_{m<j}\lambda_{j,m}(t)|j\rangle\langle m|+\mathrm{H.c.},
\end{eqnarray}
where the real number $\lambda_{j,j}(t)$ denotes level energies, and the complex number $\lambda_{j,m}(t)$ is the coupling strength between $|j\rangle$ and $|m\rangle$. Then, the expression of the propagator can be written as
\begin{eqnarray} \label{s48a}
U(t)=\sum_{j,m=1}^{J}U_{jm}|j\rangle\langle m|,
\end{eqnarray}
where the complex number $U_{jm}$ characterizes the transition between $|j\rangle$ and $|m\rangle$.

When introducing extra constant phases to the coupling strengths, the Hamiltonian becomes
\begin{eqnarray} \label{s50}
H(t,\bm{\theta})&=&\sum_{j=1}^{J}\lambda_{j,j}(t)|j\rangle\langle j|+\sum_{m<j}\lambda_{j,m}(t)e^{i\theta_{jm}}|j\rangle\langle m|+\mathrm{H.c.}
\end{eqnarray}
where $\bm{\theta}=(\theta_{12},\dots,\theta_{1J},\theta_{23}, \theta_{24},\dots)$ parameterizes different constant phases in external fields. In this situation, the total number of arbitrarily adjustable phases in external fields is
\begin{eqnarray}
(J-1)+(J-2)+\cdots+1=J(J-1)/2,
\end{eqnarray}

Next, we rewrite the Hamiltonian (\ref{s50}) in a set of new basis $|\tilde{j}\rangle$, which is connected to the original basis according to the following relations: $|\tilde{j}\rangle=e^{\theta_j}|j\rangle$, and the form reads
\begin{eqnarray} \label{s51}
H(t,\bm{\theta})&=&\sum\lambda_{j,j}(t)|\tilde{j}\rangle\langle \tilde{j}|+\sum_{m<j}\lambda_{j,m}(t)e^{i(\theta_{jm}+\theta_{j}-\theta_{m})}|\tilde{j}\rangle\langle \tilde{m}|+\mathrm{H.c.}
\end{eqnarray}
To make the form of Eq.~(\ref{s51}) be the same as Eq.~(\ref{s49a}), we demand
\begin{eqnarray} \label{s52}
\theta_{jm}+\theta_{j}-\theta_{m}=0.
\end{eqnarray}
Note that the number of $\theta_{jm}$ is $J(J-1)/2$, while it is only $J$ for $\theta_j$.
As a result, the number of the phases $\theta_{jm}$ that can take arbitrary value is not more than $J$, and the rest needs to satisfy Eq.~(\ref{s52}).
For the Hamiltonian with near-neighbor interactions, as shown in the main text, there are exactly $J$ phases in external fields. Therefore, all of them can be chosen arbitrary.

When the phases $\theta_{jm}$ in external fields satisfy Eq.~(\ref{s52}), the expression of the propagator also reads
\begin{eqnarray}
U(t)&=&\sum_{j,m=1}^{J}U_{jm}|\tilde{j}\rangle\langle \tilde{m}|, \cr
&=&\sum_{j,m=1}^{J}U_{jm}e^{i(\theta_j-\theta_n)}|j\rangle\langle m|,
\end{eqnarray}
where $U_{jm}$ exactly comes from Eq.~(\ref{s48a}).
This means that after introducing extra phases in external fields, the transition amplitude is still left unchange but different phases.

\end{widetext}

\end{document}